\documentclass[a4paper]{article}
\usepackage{amssymb,amsmath,epsfig,cite}
\usepackage{a4,times,setspace}
\usepackage{graphicx}

\newcommand{\di}{\genfrac{}{}{0pt}{}}
\newcounter{multieqs}

\newcommand{\be}{\begin{equation}}
\newcommand{\ee}{\end{equation}}
\newcommand{\eq}[1]{(\ref{#1})}

\def\bea{\begin{eqnarray}}
\def\eea{\end{eqnarray}}
\def\beb{\begin{eqnarray*}}
\def\eeb{\end{eqnarray*}}

\def\bit{\begin{itemize}}
\def\eit{\end{itemize}}

\def\one{\mbox{1 \kern-.59em {\rm l}}}

  \def\la{\lambda}

 \def\cH{{\cal H}} 
  
\def\cM{{\cal M}}

\def\R{{\mathbb R}}

\def\one{1\!\!1\,\,}

\begin{document}
\thispagestyle{empty}








\begin{flushright}
hep-th/0607208\\
UWThPh-2006-14\\
\end{flushright}

\vspace{0.3cm}
\boldmath
\begin{center}
  {\large {\bf Noncommutative QFT and Renormalization\footnote{Talk given by H.~Grosse at the 
Corfu satellite workshop "Noncommutative Geometry in Field and String Theory", 2005}}}
\end{center}
\unboldmath
\begin{center}
                                                                                                                                         
{\bf
Harald~Grosse\footnote{Email: harald.grosse@univie.ac.at}
and
Michael~Wohlgenannt\footnote{Supported by the Austrian Science 
Fund, project P18657-N16. Email: michael.wohlgenannt@univie.ac.at}
}                                                                                                                                         
\end{center}
\begin{center}
Universit\"at Wien, Institut f\"ur Theoretische Physik, \\
  Boltzmanngasse 5, A-1090 Wien, Austria\\
\vspace{.3cm}
\end{center}
\vskip 1em

\begin{abstract}
Field theories on deformed spaces suffer from the IR/UV mixing and
renormalization is generically spoiled. In work with R.~Wulkenhaar, one
of us realized a way to cure this disease by adding one more marginal 
operator. We review these ideas, show the application to $\phi^3$ models and use
the heat kernel expansion methods for a scalar field theory coupled
to an external gauge field on a $\theta$-deformed space and derive
noncommutative gauge field actions. 
\end{abstract}


\section{\bf Introduction}

Four-dimensional quantum field theory suffers from infrared and
ultraviolet divergences as well as from the divergence of the
renormalized perturbation expansion. Despite the impressive
agreement between theory and experiments and despite many
attempts, these problems are not settled and remain a big
challenge for theoretical physics. Furthermore, attempts to
formulate a quantum theory of gravity have not yet been fully
successful. It is astonishing that the two pillars of modern
physics, quantum field theory and general relativity, seem to be
incompatible. This convinced physicists to look for more general
descriptions: After the formulation of supersymmetry and supergravity, string theory was
developed, and anomaly cancellation forced the introduction of six
additional dimensions. On the other hand, loop gravity was
formulated, and led to spin networks and space-time foams. Both
approaches are not fully satisfactory. A third impulse came from
noncommutative geometry developed by Alain Connes, providing a
natural interpretation of the Higgs effect at the classical level.
This finally led to noncommutative quantum field theory, which is
the subject of this contribution. It allows to incorporate
fluctuations of space into quantum field theory. There are of
course relations among these three developments. In particular,
the field theory limit of string theory leads to certain
noncommutative field theory models, and some models defined over
fuzzy spaces are related to spin networks.

The argument that space-time should be modified at very short
distances goes back to Schr\"odinger and Heisenberg.
Noncommutative coordinates appeared already in the work of Peierls
for the magnetic field problem, and are obtained after projecting
onto a particular Landau level. Pauli communicated this to
Oppenheimer, whose student Snyder \cite{Snyder:1947qz} wrote down
the first deformed space-time algebra preserving Lorentz symmetry.
After the development of noncommutative geometry by Connes
\cite{Connes:1986uh}, it was first applied in physics to the
integer quantum Hall effect. Gauge models on the two-dimensional
noncommutative tori were formulated, and the relevant projective
modules over this space were classified.

Through interactions
with John Madore one of us (H.G.) realized that such Fuzzy geometries allow to
obtain natural cutoffs for quantum field theory
\cite{Grosse:1992bm}. This line of work was further developed
together with Peter Pre\v{s}najder and Ctirad Klim\v{c}\'{\i}k
\cite{Grosse:1995ar}. At almost the same time, Filk
\cite{Filk:1996dm} developed his Feynman rules for the canonically
deformed four-dimensional field theory, and Doplicher, Fredenhagen
and Roberts \cite{Doplicher:1994tu} published their work on
deformed spaces. The subject experienced a major boost after one
realized that string theory leads to noncommutative field theory
under certain conditions \cite{Schomerus:1999ug,Seiberg:1999vs},
and the subject developed very rapidly; see e.g.
\cite{Szabo:2001kg,Douglas:2001ba}.

\section{Noncommutative Quantum Field Theory}

The formulation of  Noncommutative Quantum Field Theory (NCFT)
follows a dictionary worked out by
mathematicians. Starting from some manifold $\cM$ one obtains
the commutative algebra of smooth
functions over $\cM$, which is then quantized along with
additional structure. Space itself then looks locally like a
phase space in quantum mechanics.
Fields are elements of the algebra respectively
a finitely generated projective module, and
integration is replaced by a suitable trace operation.



Following these lines, one obtains field theory on quantized (or
deformed) spaces, and Feynman rules for a perturbative expansion
can be worked out. However some unexpected features such as IR/UV
mixing arise upon quantization, which are described below. In 2000
Minwalla, van Raamsdonk and Seiberg realized
\cite{Minwalla:1999px} that  perturbation theory for field
theories defined on the Moyal plane faces a serious problem. The
planar contributions show the standard singularities which can be
handled by a renormalization procedure. The nonplanar one loop
contributions are finite for generic momenta, however they become
singular at exceptional momenta. The usual UV divergences are then
reflected in new singularities in the infrared, which is called
IR/UV mixing. This spoils the usual renormalization procedure:
Inserting many such loops to a higher order diagram generates
singularities of any inverse power. Without imposing a special
structure such as supersymmetry, the renormalizability seems lost;
see also \cite{Chepelev:1999tt,Chepelev:2000hm}.

However, progress was made recently, when H.G. and R.~Wulkenhaar were able to give a
solution of this problem for the special case of a scalar
four-dimensional theory defined on the Moyal-deformed space
$\R^4_\theta$ \cite{Grosse:2004yu}. The IR/UV mixing contributions
were taken into account through a modification of the free Lagrangian
by adding an oscillator term with parameter $\Omega$, which modifies
the spectrum of the free Hamiltonian. The harmonic oscillator term was
obtained as a result of the renormalization proof. The model fulfills
then the Langmann-Szabo duality \cite{Langmann:2002cc} relating short
distance and long distance behavior. The proof follows ideas of
Polchinski. There are indications that a constructive procedure might
be possible and give a nontrivial $\phi^4$ model, which is currently
under investigation \cite{Rivasseau:2005bh}. At $\Omega =1$ the model
becomes self-dual, and we are presently studying them in more
detail. The noncommutative Euclidean 
selfdual $\phi^3$ model can be solved using the relationship to the Kontsevich
matrix model. This relation holds for any even dimension, but a renormalization
still has to be applied. In $D=2$ and $D=4$ dimensions
the models are super-renormalizable \cite{Grosse:2005ig,Grosse:2006qv}. 
In $D=6$ dimensions, the model 
is only renormalizable and details are presently worked out \cite{Grosse:2006hs}.

Nonperturbative aspects of NCFT have also been studied in recent
years.  The most significant and surprising result is that the
IR/UV mixing can lead to a new phase denoted as ``striped phase''
\cite{Gubser:2000cd}, where translational symmetry is
spontaneously broken. The existence of such a phase has indeed
been confirmed in numerical studies
\cite{Bietenholz:2004xs,Martin:2004un}. To understand better the
properties of this phase and the phase transitions, further work
and better analytical techniques are required, combining results
from perturbative renormalization with nonperturbative techniques.
Here a particular feature of scalar NCFT is very suggestive: the
field can be described as a hermitian matrix, and the quantization
is defined nonperturbatively by integrating over all such
matrices. This provides a natural starting point for
nonperturbative studies. In particular, it suggests and allows to
apply ideas and techniques from random matrix theory.

Remarkably, gauge theories on quantized spaces can also be formulated
in a similar way \cite{Ambjorn:2000nb,Carow-Watamura:1998jn,Steinacker:2003sd,Behr:2005wp}.
The action can be written as multi-matrix
models, where the gauge fields are encoded in terms of matrices which
can be interpreted as ``covariant coordinates''. The field strength
can be written as commutator, which induces the usual
kinetic terms in the commutative limit. Again, this allows a natural
nonperturbative quantization in terms of matrix integrals.

In the last section, we discuss a formulation of gauge theories related to 
the approach to NCFT presented here. We start with noncommutative $\phi^4$ theory
on canonically deformed Euclidean space with additional oscillator potential.
The oscillator potential modifies the free theory and solves the IR/UV mixing
problem. We couple an external gauge field to the scalar field via introducing covariant
coordinates. As in the classical case, we extract the dynamics of the gauge field
from the divergent contributions to the 1-loop effective action. The effective
action is calculated using a heat kernel expansion \cite{Gilkey:1995mj,Vassilevich:2003xt}.
The technical details are going to be presented in \cite{Grosse:2006mw}.

\section{Renormalization of $\phi^4$-theory 
on the $4D$ Moyal plane}

We briefly sketch the methods used in 
\cite{Grosse:2004yu}  proving the renormalizability for scalar
field theory defined on the 4-dimensional quantum plane
$\R^4_\theta$, with commutation relations 
\be
\label{CCR}
[x_\mu,x_\nu] = i \theta_{\mu\nu}\,. 
\ee
The IR/UV mixing was taken into account
through a modification of the free Lagrangian, by adding an
oscillator term which modifies the spectrum of the free
Hamiltonian: \be S= \int d^4x \Big( \frac{1}{2} \partial_\mu \phi
\star
\partial^\mu \phi + \frac{\Omega^2}{2} (\tilde{x}_\mu \phi) \star
(\tilde{x}^\mu \phi) + \frac{\mu^2}{2} \phi \star \phi +
\frac{\lambda}{4!} \phi \star \phi \star \phi \star
\phi\Big)(x)\;. \label{action}
\ee
Here, $\tilde{x}_\mu=2(\theta^{-1})_{\mu\nu} x^\nu$ and 
$\star$ is the Moyal star product
\begin{align}
  (a\star b)(x):= \int d^4y \frac{d^4k}{(2\pi)^4}
  a(x{+}\tfrac{1}{2}\theta {\cdot} k) b(x{+}y)
  \,\mathrm{e}^{\mathrm{i}ky}\;, \qquad
  \theta_{\mu\nu}=-\theta_{\nu\mu} \in \mathbb{R}\;.
\end{align}
The model is covariant under the
Langmann-Szabo duality relating short distance and long distance
behavior. At $\Omega =1$ the model becomes self-dual, and
connected to integrable models. 

The renormalization proof proceeds by using a matrix base,
which leads to a dynamical matrix model of the type:
\be
S[\phi]  =(2\pi\theta)^2 \sum_{m,n,k,l\in \mathbb{N}^2}
\Big(\dfrac{1}{2} \phi_{mn} \Delta_{mn;kl} \phi_{kl} +
\frac{\lambda}{4!} \phi_{mn}\phi_{nk} \phi_{kl} \phi_{lm}\Big)\;,
\label{Sm}
\ee
where
\begin{align}
\Delta_{\di{m^1}{m^2}\di{n^1}{n^2};\di{k^1}{k^2}\di{l^1}{l^2}} &=
\big(\mu^2{+} \tfrac{2{+}2\Omega^2}{\theta}
(m^1{+}n^1{+}m^2{+}n^2{+}2) \big) \delta_{n^1k^1} \delta_{m^1l^1}
\delta_{n^2k^2} \delta_{m^2l^2} \nonumber
\\
&- \tfrac{2{-}2\Omega^2}{\theta} \big(\sqrt{k^1l^1}\,
  \delta_{n^1+1,k^1}\delta_{m^1+1,l^1} + \sqrt{m^1n^1}\,
  \delta_{n^1-1,k^1} \delta_{m^1-1,l^1}\big)\delta_{n^2k^2}
  \delta_{m^2l^2}
\nonumber
\\
&- \tfrac{2{-}2\Omega^2}{\theta} \big(\sqrt{k^2l^2}\,
  \delta_{n^2+1,k^2}\delta_{m^2+1,l^2} + \sqrt{m^2n^2}\,
  \delta_{n^2-1,k^2} \delta_{m^2-1,l^2}\big)\delta_{n^1k^1}
  \delta_{m^1l^1} \;.
\label{Gm}
\end{align}
The interaction part becomes a trace of product of matrices, and no
oscillations occur in this basis. The propagator obtained from the
free part is quite complicated, in 4  dimensions it is:
\begin{align}
&G_{\di{m^1}{m^2}\di{n^1}{n^2}; \di{k^1}{k^2}\di{l^1}{l^2}}
\nonumber
\\*
&= \frac{\theta}{2(1{+}\Omega)^2} \!
\sum_{v^1=\frac{|m^1-l^1|}{2}}^{\frac{m^1+l^1}{2}}
\sum_{v^2=\frac{|m^2-l^2|}{2}}^{\frac{m^2+l^2}{2}} \!\! B\big(1{+}
\tfrac{\mu^2 \theta}{8\Omega}
 {+}\tfrac{1}{2}(m^1{+}k^1{+}m^2{+}k^2){-}v^1{-}v^2,
1{+}2v^1{+}2v^2 \big) \nonumber
 \\
&\times {}_2F_1\bigg(\di{1{+} 2v^1{+}2v^2\,,\; \frac{\mu^2
\theta}{8\Omega}
 {-}\frac{1}{2}(m^1{+}k^1{+}m^2{+}k^2){+}v^1{+}v^2
}{2{+} \frac{\mu^2 \theta}{8\Omega}
 {+}\frac{1}{2}(m^1{+}k^1{+}m^2{+}k^2){+}v^1{+}v^2} \bigg|
\frac{(1{-} \Omega)^2}{(1{+}\Omega)^2}\bigg) \Big(\frac{1{-}
\Omega}{1{+}\Omega}\Big)^{2v^1+2v^2} \nonumber
 \\
&\times \prod_{i=1}^2 \delta_{m^i+k^i,n^i+l^i} \sqrt{
\binom{n^i}{v^i{+}\frac{n^i-k^i}{2}}
\binom{k^i}{v^i{+}\frac{k^i-n^i}{2}}
\binom{m^i}{v^i{+}\frac{m^i-l^i}{2}}
\binom{l^i}{v^i{+}\frac{l^i-m^i}{2}}}\;. \label{prop}
\end{align}
These propagators (in 2 and 4 dimensions) show asymmetric decay
properties: 
\begin{align}
\parbox{130mm}{\begin{picture}(120,40)
\put(-25,-105){\epsfig{file=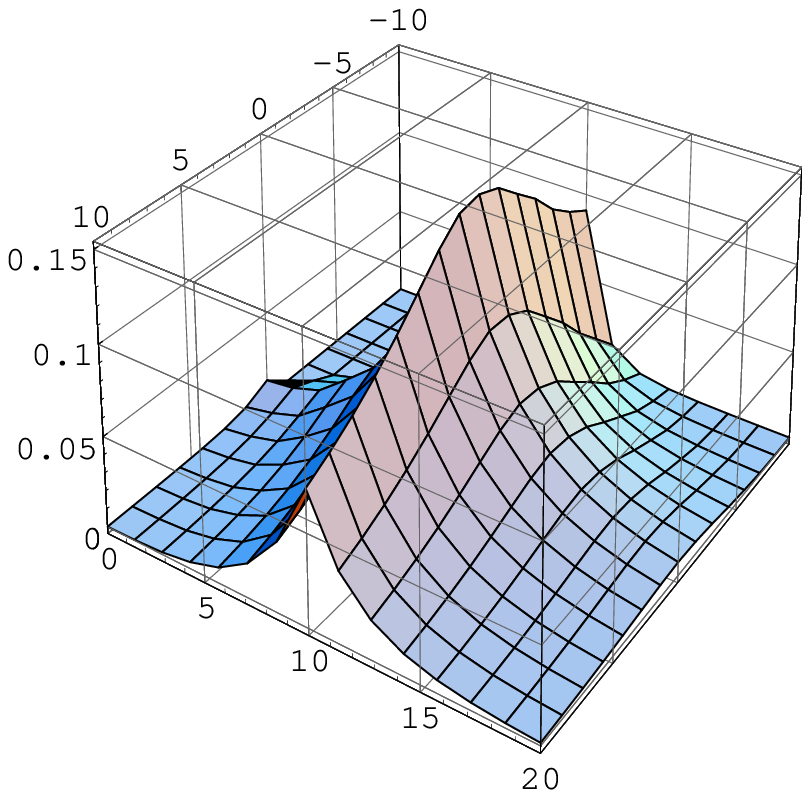,scale=0.55,bb=0 0 598 843}}
\put(80,10){\vector(0,1){20}} \put(80,10){\vector(4,-1){15}}
\put(80,10){\vector(3,1){10}}
\put(64,33){\mbox{\footnotesize$\theta^{-1}
\Delta_{\di{10}{0}\di{10+\alpha}{0};\di{l+\alpha}{0} \di{l}{0}}$}}
\put(92,15){\mbox{\footnotesize$\alpha$}}
\put(97,5){\mbox{\footnotesize$l$}}
\put(40,0){\mbox{\footnotesize$\Omega=0.1$}}
\put(55,0){\mbox{\footnotesize$\mu=0$}}
\end{picture}}
\label{quasi-local-1}
\end{align}
They decay exponentially on particular directions (in $l$-direction in
the picture), but have power law decay in others (in
$\alpha$-direction in the picture). These decay properties are crucial
for the perturbative renormalizability of the models.

The proof in \cite{Grosse:2003aj,Grosse:2004yu} follows the ideas of
Polchinski \cite{Polchinski:1983gv}. The quantum field theory
corresponding to the action (\ref{Sm}) is defined --- as usual --- by
the partition function 
\begin{align}
Z[J]  = \int \left(\prod_{m,n} d\phi_{mn}\right) \;\exp\left(-S[\phi]-
\sum_{m,n} \phi_{mn} J_{nm}\right)\;.
\end{align}
The strategy due to Wilson \cite{Wilson:1973jj} consists in integrating in
the first step only those field modes $\phi_{mn}$ which have a matrix index
bigger than some scale $\theta\Lambda^2$. The result is an effective
action for the remaining field modes which depends on $\Lambda$. One
can now adopt a smooth transition between integrated and not
integrated field modes so that the $\Lambda$-dependence of the
effective action is given by a certain differential equation, the
Polchinski equation. 

Now, renormalization amounts to prove that the Polchinski equation
admits a regular solution for the effective action which depends on
only a finite number of initial data. This requirement is hard to
satisfy because the space of effective actions is infinite dimensional
and as such develops an infinite dimensional space of singularities
when starting from generic initial data.  

The Polchinski equation can be iteratively solved in perturbation
theory where it can be graphically written as  
\begin{align}
\Lambda \frac{\partial}{\partial \Lambda}
&\parbox{26mm}{\begin{picture}(20,21)
 \put(0,0){\epsfig{file=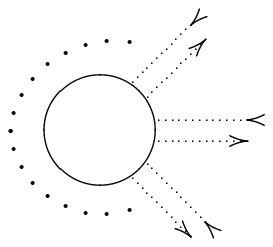,bb=69 621 141 684}}
 \put(22,13.5){\mbox{\scriptsize$n_1$}}
 \put(20,8){\mbox{\scriptsize$m_1$}}
 \put(21,3){\mbox{\scriptsize$n_2$}}
 \put(13,1){\mbox{\scriptsize$m_2$}}
 \put(19,17){\mbox{\scriptsize$m_N$}}
 \put(15,23){\mbox{\scriptsize$n_N$}}
\end{picture}}
\nonumber
\\*[-4ex]
& =\frac{1}{2}\sum_{m,n,k,l} \sum_{N_1=1}^{N-1}
\parbox{48mm}{\begin{picture}(48,25)
 \put(0,0){\epsfig{file=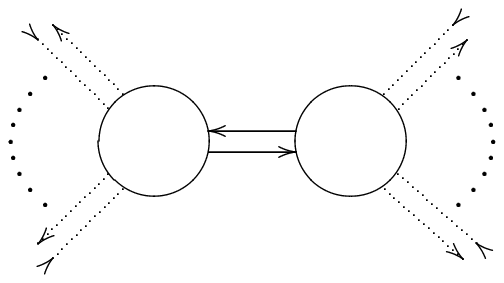,bb=69 615 214 684}}
 \put(0,4){\mbox{\scriptsize$m_1$}}
 \put(6,0){\mbox{\scriptsize$n_1$}}
 \put(0,21){\mbox{\scriptsize$n_{N_1}$}}
 \put(7,23){\mbox{\scriptsize$m_{N_1}$}}
 \put(47,20){\mbox{\scriptsize$m_{N_1+1}$}}
 \put(37,26){\mbox{\scriptsize$n_{N_1+1}$}}
 \put(48,3){\mbox{\scriptsize$n_N$}}
 \put(39,2){\mbox{\scriptsize$m_N$}}
 \put(27,14.5){\mbox{\scriptsize$k$}}
 \put(28,8){\mbox{\scriptsize$l$}}
 \put(22,15){\mbox{\scriptsize$n$}}
 \put(22,9){\mbox{\scriptsize$m$}}
\end{picture}}
\quad -  \frac{1}{4\pi \theta} \sum_{m,n,k,l}
\parbox{35mm}{\begin{picture}(30,35)
 \put(0,0){\epsfig{file=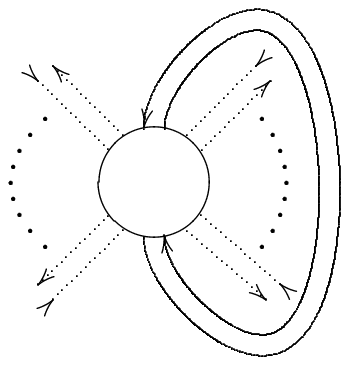,bb=69 585 168 687}}
 \put(0,9){\mbox{\scriptsize$m_1$}}
 \put(6,5){\mbox{\scriptsize$n_1$}}
 \put(-1,26){\mbox{\scriptsize$n_{i-1}$}}
 \put(7.5,29){\mbox{\scriptsize$m_{i-1}$}}
 \put(27,25){\mbox{\scriptsize$m_i$}}
 \put(22.5,30){\mbox{\scriptsize$n_i$}}
 \put(27,9){\mbox{\scriptsize$n_N$}}
 \put(20,7){\mbox{\scriptsize$m_N$}}
 \put(12,25.5){\mbox{\scriptsize$n$}}
 \put(18,26){\mbox{\scriptsize$m$}}
 \put(12,10.5){\mbox{\scriptsize$k$}}
 \put(18,10.5){\mbox{\scriptsize$l$}}
\end{picture}}
\label{Lgraph}
\end{align}
The graphs are graded by the number of vertices and the number of
external legs. Then, to the $\Lambda$-variation of a graph on the lhs
there only contribute graphs with a smaller number of vertices and a
bigger number of legs. A general graph is thus obtained by iteratively
adding a propagator to smaller building blocks, starting with the
initial $\phi^4$-vertex, and integrating over $\Lambda$. Here, these
propagators are differentiated cut-off propagators
$Q_{mn;kl}(\Lambda)$ which vanish (for an appropriate choice of the
cut-off function) unless the maximal index is in the interval
$[\theta\Lambda^2,2\theta \Lambda^2]$. As the field carry two matrix
indices and the propagator four of them, the graphs are ribbon graphs
familiar from matrix models.

It can then be shown that cut-off propagator $Q(\Lambda)$ is
bounded by $\frac{C}{\theta \Lambda^2}$. This was achieved numerically
in \cite{Grosse:2004yu} and later confirmed analytically in
\cite{Rivasseau:2005bh}. A nonvanishing frequency parameter $\Omega$
is required for such a decay behavior. As the volume of each
two-component index $m\in \mathbb{N}^2$ is bounded by $C' \theta^2
\Lambda^4$ in graphs of the above type, the power counting degree of
divergence is (at first sight) $\omega=4S-2I$, where $I$ is the number
of propagators and $S$ the number of summation indices.

It is now important to take into account that if three indices of a
propagator $Q_{mn;kl}(\Lambda)$ are given, the fourth one is
determined by $m+k=n+l$, see (\ref{prop}). Then, for simple planar
graphs one finds that $\omega=4-N$ where $N$ is the number of external
legs. But this conclusion is too early, there is a difficulty in
presence of completely inner vertices, which require additional index
summations. The graph
\begin{align}
\parbox{50\unitlength}{\begin{picture}(50,32)
       \put(0,0){\epsfig{file=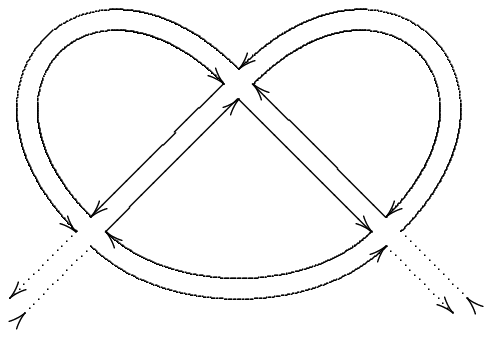,bb=92 571 226 661}}
       \put(0,5){\mbox{\scriptsize$m$}}
       \put(3,0){\mbox{\scriptsize$n$}}
       \put(44,6){\mbox{\scriptsize$l$}}
       \put(40,2){\mbox{\scriptsize$k$}}
       \put(24,29){\mbox{\scriptsize$q$}}
       \put(6,15){\mbox{\scriptsize$p_1{+}m$}}
       \put(14,23){\mbox{\scriptsize$p_1{+}q$}}
       \put(35,15){\mbox{\scriptsize$p_2{+}l$}}
       \put(27,23){\mbox{\scriptsize$p_2{+}q$}}
       \put(20,17){\mbox{\scriptsize$p_3{+}q$}}
       \put(29,8){\mbox{\scriptsize$p_3{+}l$}}
       \put(12,9){\mbox{\scriptsize$p_3{+}m$}}
\end{picture}}
\label{graph-at2}
\end{align}
entails four independent summation indices $p_1,p_2,p_3$ and $q$,
whereas for the powercounting degree $2=4-N=4S-5\cdot 2$ we should
only have $S=3$ of them. It turns out that due to the quasi-locality
of the propagator (the exponential decay in $l$-direction in
(\ref{quasi-local-1})), the sum over $q$ for fixed $m$ can be
estimated without the need of the volume factor.

Remarkably, the quasi-locality of the propagator not only ensures the
correct powercounting degree for planar graphs, it also renders all
nonplanar graphs superficially convergent. 
For instance, in the nonplanar graphs
\begin{align}
&\left.\parbox{43mm}{\begin{picture}(20,20)
       \put(0,0){\epsfig{file=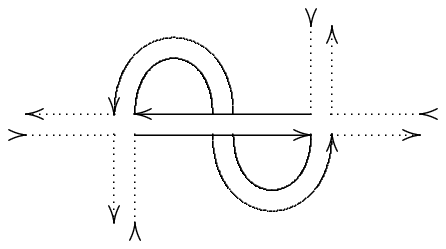,bb=71 625 187 684}}
       \put(2,13){\mbox{\scriptsize$m_4$}}
       \put(0,7){\mbox{\scriptsize$n_4$}}
       \put(4,2){\mbox{\scriptsize$m_1$}}
       \put(13,0){\mbox{\scriptsize$n_1$}}
       \put(36,13){\mbox{\scriptsize$n_2$}}
       \put(34,7){\mbox{\scriptsize$m_2$}}
       \put(32,18){\mbox{\scriptsize$m_3$}}
       \put(25,20){\mbox{\scriptsize$n_3$}}
       \put(13.5,13.5){\mbox{\scriptsize$q$}}
       \put(25,6.5){\mbox{\scriptsize$q'$}}
   \end{picture}} \right|_{q'=n_1+n_3-q}~
\left.\parbox{41mm}{\begin{picture}(20,24)
       \put(0,0){\epsfig{file=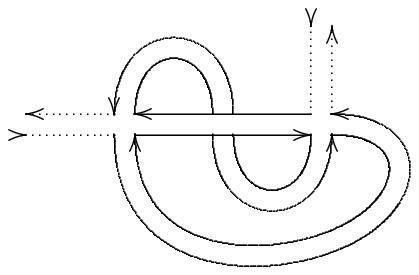,bb=71 613 184 684}}
       \put(2,17){\mbox{\scriptsize$m_2$}}
       \put(0,11){\mbox{\scriptsize$n_2$}}
       \put(13,11){\mbox{\scriptsize$r'$}}
       \put(32,11){\mbox{\scriptsize$r$}}
       \put(32,22){\mbox{\scriptsize$m_1$}}
       \put(25,24){\mbox{\scriptsize$n_1$}}
       \put(13.5,17.5){\mbox{\scriptsize$q$}}
       \put(25,10.5){\mbox{\scriptsize$q'$}}
   \end{picture}} \right|_{\mbox{\scriptsize$\begin{array}{l}
q'=m_2+r-q \\ r'=n_2+r-m_1 \end{array}$}} \hspace*{-1em}
\label{np-graphs}
\end{align}
the summation over $q$ and $q,r$, respectively, is of the same type as
over $q$ in (\ref{graph-at2}) so that the graphs in (\ref{np-graphs})
can be estimated without any volume factor.

After all, we have obtained the powercounting degree of divergence
\begin{align}
\omega =4-N -4(2g+B-1)  
\label{pcd}
\end{align}
for a general ribbon graph, where $g$ is the genus of the Riemann
surface on which the graph is drawn and $B$ the number of holes in the
Riemann surface. Both are directly determined by the graph. It should
be stressed, however, that although the number (\ref{pcd}) follows
from counting the required volume factors, its proof in our scheme is
not so obvious: The procedure consists of adding a new cut-off
propagator to a given graph, and in doing so the topology $(B,g)$ has
many possibilities to arise from the topologies of the smaller parts
for which one has estimates by induction. The proof that in every
situation of adding a new propagator one obtains (\ref{pcd}) 
is given in \cite{Grosse:2003aj}. Moreover, the boundary
conditions for the integration have to be correctly chosen to confirm
(\ref{pcd}), see below.

The powercounting behavior (\ref{pcd}) is good news because it
implies that (in contrast to the situation without the oscillator
potential) all nonplanar graphs are superficially convergent. 
However, this does not mean that all problems are solved: The
remaining planar two- and four-leg graphs which are divergent carry
matrix indices, and (\ref{pcd}) suggests that these are divergent
independent of the matrix indices. An infinite number of adjusted
initial data would be necessary in order to remove these divergences.

Fortunately, a more careful analysis shows that the powercounting
behavior is improved by the index jump along the trajectories of the
graph. For example, the index jump for the graph (\ref{graph-at2}) is
defined as $J=\|k-n\|_1+\|q-l\|_1+\|m-q\|_1$. Then, the amplitude is
suppressed by a factor of order $\left(\dfrac{\max(m,n\dots)}{\theta
    \Lambda^2}\right)^{\frac{J}{2}}$ compared with the naive
estimation. Thus, only planar four-leg graphs with $J=0$ and planar
two-leg graphs with $J=0$ or $J=2$ are divergent (the total jumps is
even). For these cases, a discrete Taylor expansion
about the graphs with vanishing indices is employed. Only the leading terms of the
expansion, i.e.\ the reference graphs with vanishing indices, are
divergent whereas the difference between original graph and reference
graph is convergent. Accordingly, in this scheme only the reference
graphs must be integrated in a way that involves initial conditions. 
For example, if the contribution to the rhs of the Polchinski equation
(\ref{Lgraph}) is given by the graph 
\begin{align}
\Lambda \frac{\partial}{\partial \Lambda}
A^{(2)\text{planar,1PI}}_{mn;nk;kl;lm}[\Lambda] = \sum_{p \in
  \mathbb{N}^2} \left(\quad
  \parbox{38mm}{\begin{picture}(38,15)
       \put(0,0){\epsfig{file=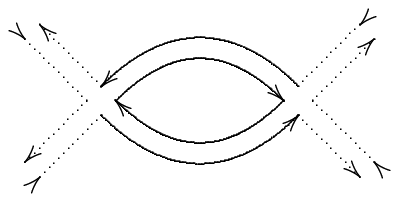,bb=71 630 174 676}}
       \put(-4,12){\mbox{\footnotesize$m$}}
       \put(-1,6){\mbox{\footnotesize$m$}}
       \put(34,9){\mbox{\footnotesize$k$}}
       \put(36,4){\mbox{\footnotesize$k$}}
       \put(4,-1){\mbox{\footnotesize$n$}}
       \put(30,-1){\mbox{\footnotesize$n$}}
       \put(5,14){\mbox{\footnotesize$l$}}
       \put(30,15){\mbox{\footnotesize$l$}}
       \put(12,8){\mbox{\footnotesize$p$}}
       \put(22,8){\mbox{\footnotesize$p$}}
   \end{picture}} \right)(\Lambda) \;,
\label{A4diff}
\end{align}
the $\Lambda$-integration is performed as follows:
\begin{align}
&A^{(2)\text{planar,1PI}}_{mn;nk;kl;lm}[\Lambda] \nonumber
\\*[-1ex]
&\quad =  -\int_{\Lambda}^{\infty} \frac{d \Lambda'}{\Lambda'}
\, \sum_{p \in \mathbb{N}^2} \left(~~~
\parbox{39mm}{\begin{picture}(20,15)
       \put(0,0){\epsfig{file=a24,bb=71 630 174 676}}
       \put(-4,12){\mbox{\footnotesize$m$}}
       \put(-1,6){\mbox{\footnotesize$m$}}
       \put(34,9){\mbox{\footnotesize$k$}}
       \put(36,4){\mbox{\footnotesize$k$}}
       \put(4,-1){\mbox{\footnotesize$n$}}
       \put(30,-1){\mbox{\footnotesize$n$}}
       \put(5,14){\mbox{\footnotesize$l$}}
       \put(30,15){\mbox{\footnotesize$l$}}
       \put(12,8){\mbox{\footnotesize$p$}}
       \put(22,8){\mbox{\footnotesize$p$}}
   \end{picture}}
-~~
\parbox{40mm}{\begin{picture}(20,15)
       \put(0,0){\epsfig{file=a24,bb=71 630 174 676}}
       \put(-4,12){\mbox{\footnotesize$m$}}
       \put(-1,6){\mbox{\footnotesize$m$}}
       \put(34,9){\mbox{\footnotesize$k$}}
       \put(36,4){\mbox{\footnotesize$k$}}
       \put(4,-1){\mbox{\footnotesize$n$}}
       \put(30,-1){\mbox{\footnotesize$n$}}
       \put(5,14){\mbox{\footnotesize$l$}}
       \put(30,15){\mbox{\footnotesize$l$}}
       \put(8,12){\mbox{\footnotesize$0$}}
       \put(25.5,12){\mbox{\footnotesize$0$}}
       \put(8,2){\mbox{\footnotesize$0$}}
       \put(25.5,2){\mbox{\footnotesize$0$}}
       \put(12,8){\mbox{\footnotesize$p$}}
       \put(22,8){\mbox{\footnotesize$p$}}
   \end{picture}}
\right)\![\Lambda'] \nonumber
\\*[2ex]
&\quad + ~~\parbox{20mm}{\begin{picture}(20,15)
       \put(0,0){\epsfig{file=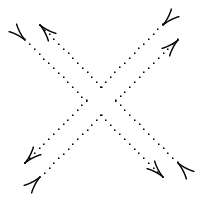,bb=71 638 117 684}}
   \put(-4,12){\mbox{\footnotesize$m$}}
       \put(-1,6){\mbox{\footnotesize$m$}}
       \put(13,9){\mbox{\footnotesize$k$}}
       \put(14,4){\mbox{\footnotesize$k$}}
       \put(4,-1){\mbox{\footnotesize$n$}}
       \put(10,-1){\mbox{\footnotesize$n$}}
       \put(5,14){\mbox{\footnotesize$l$}}
       \put(10,15){\mbox{\footnotesize$l$}}
   \end{picture}}  \left[
\int_{\Lambda_R}^\Lambda \frac{d \Lambda'}{\Lambda'} \, \sum_{p
\in \mathbb{N}^2} \left(~~\parbox{40mm}{\begin{picture}(20,15)
       \put(0,0){\epsfig{file=a24,bb=71 630 174 676}}
       \put(-2,11){\mbox{\footnotesize$0$}}
       \put(-1,5){\mbox{\footnotesize$0$}}
       \put(34,9){\mbox{\footnotesize$0$}}
       \put(36,4){\mbox{\footnotesize$0$}}
       \put(4,-1){\mbox{\footnotesize$0$}}
       \put(29,-1){\mbox{\footnotesize$0$}}
       \put(5,14){\mbox{\footnotesize$0$}}
       \put(30,15){\mbox{\footnotesize$0$}}
       \put(12,8){\mbox{\footnotesize$p$}}
       \put(22,8){\mbox{\footnotesize$p$}}
   \end{picture}}\right)\![\Lambda']
+ A^{(2,1,0)\text{1PI}}_{00;00;00;00}[\Lambda_R] \right]\,.
\label{A4}
\end{align}
Only one initial condition, 
$A^{(2,1,0)\text{1PI}}_{00;00;00;00}[\Lambda_R]$,
is required for an infinite number of planar four-leg graphs 
(distinguished by the matrix indices). We need one further initial
condition for the two-leg graphs with $J=2$ and two more 
initial condition for the two-leg graphs with $J=0$ (for the leading
quadratic and the subleading logarithmic divergence). This is one
condition more than in a commutative $\phi^4$-theory, and this
additional condition 
justifies a posteriori our starting point of adding one new term
to the action \eq{action}, the oscillator term $\Omega$.

\bigskip

Knowing the relevant/marginal couplings, we can compute Feynman graphs
with sharp matrix cut-off $\mathcal{N}$. The most important question
concerns the $\beta$-function appearing in the renormalisation group
equation which describes the cut-off dependence of the expansion
coefficients $\Gamma_{m_1n_1;\dots;m_Nn_N}$ of the effective action
when imposing normalisation conditions for the relevant and marginal
couplings. We have \cite{Grosse:2004by}
\begin{align}
\lim_{\mathcal{N}\to \infty} 
\Big(\mathcal{N} \frac{\partial}{\partial \mathcal{N}} 
+ N \gamma + \mu_0^2 \beta_{\mu_0} \frac{\partial}{\partial \mu_0^2} 
+ \beta_\lambda \frac{\partial}{\partial \lambda} 
+ \beta_\Omega \frac{\partial}{\partial \Omega} \Big) 
\Gamma_{m_1n_1;\dots;m_Nn_N}[\mu_0,\lambda,\Omega,\mathcal{N}] = 0\;,
\end{align}
where
\begin{align}
\beta_\lambda &=  \mathcal{N} 
\frac{\partial }{\partial \mathcal{N}}   
\Big( \lambda[\mu_{\text{phys}},\lambda_{\text{phys}},
\Omega_{\text{phys}},\mathcal{N} ]\Big) \;, &
\beta_\Omega &=  
 \mathcal{N} \frac{\partial }{\partial \mathcal{N}} 
\Big( \Omega[\mu_{\text{phys}},\lambda_{\text{phys}},
\Omega_{\text{phys}},\mathcal{N} ]\Big) \;,
\nonumber
\\
\beta_{\mu_0} &=  \frac{\mathcal{N} }{\mu_0^2}  
\frac{\partial }{\partial \mathcal{N}} 
\Big( \mu_0^2[\mu_{\text{phys}},\lambda_{\text{phys}},
\Omega_{\text{phys}},\mathcal{N} ]\Big) \;, &
\gamma &= \mathcal{N} \frac{\partial }{\partial \mathcal{N}}
\Big( \ln \mathcal{Z}[\mu_{\text{phys}},\lambda_{\text{phys}},
\Omega_{\text{phys}},\mathcal{N} ]\Big) \;.
\end{align}
Here, $\mathcal{Z}$ is the wavefunction renormalisation. To one-loop
order one finds \cite{Grosse:2004by}
\begin{align}
\beta_\lambda &=  \frac{\lambda_{\text{phys}}^2}{48 \pi^2}
\frac{(1{-}\Omega_{\text{phys}}^2)}{(1{+}\Omega_{\text{phys}}^2)^3}
\;, & \beta_\Omega &= \frac{\lambda_{\text{phys}}
\Omega_{\text{phys}}}{96 \pi^2}
\frac{(1{-}\Omega_{\text{phys}}^2)}{(1{+}\Omega_{\text{phys}}^2)^3}\;,
\label{beta}
\\*
\beta_{\mu} &= - \dfrac{\lambda_{\text{phys}}
\Big(4\mathcal{N}\ln(2) + \frac{(8{+}
\theta\mu_{\text{phys}}^2)\Omega^2_{\text{phys}}}{
(1{+}\Omega_{\text{phys}}^2)^2} \Big)}{48 \pi^2 \theta
  \mu_{\text{phys}}^2 (1{+}\Omega_{\text{phys}}^2) }\;,
& \gamma &=  \frac{\lambda_{\text{phys}} }{96 \pi^2}
\frac{\Omega^2_{\text{phys}}}{(1{+}\Omega_{\text{phys}}^2)^3}\;.
\end{align}
Eq.~(\ref{beta}) shows that the ratio
of the coupling constants $\frac{\lambda}{\Omega^2}$ remains bounded along
the renormalization group flow up to first order. 
Starting from given small values
for $\Omega_R,\lambda_R$ at $\mathcal{N}_R$, the frequency grows
in a small region around $\ln
\frac{\mathcal{N}}{\mathcal{N}_R}=\frac{48\pi^2}{\lambda_R}$ to
$\Omega \approx 1$. The coupling constant approaches
$\lambda_\infty=\frac{\lambda_R}{\Omega_R^2}$, which can be made
small for sufficiently small $\lambda_R$. This leaves the chance
of a nonperturbative construction \cite{Rivasseau:1991ub} of the
model.

In particular, the $\beta$-function vanishes at the self-dual point
$\Omega = 1$, indicating special properties of the model.

\section{Nontrivial solvable $\phi^3$ model} \label{subsec:MM-techniques}

In \cite{Grosse:2006qv} the 4-dimensional scalar noncommutative 
$\phi^3$ model is considered, with additional oscillator-type potential 
in order to avoid the problem of IR/UV mixing. 
The model is defined by the action \cite{Grosse:2005ig,Grosse:2006qv}
\be 
\tilde S = \int_{\R^{4}_{\theta}} \frac 12 \partial_i\phi
\partial_i\phi + \frac {\mu^2}2 \phi^2 + \Omega^2 (\tilde x_i \phi)
(\tilde x_i \phi) + \frac{i\tilde\lambda}{3!}\;\phi^3 
\ee 
on the $4$-dimensional quantum plane.
The dynamical object is the scalar field
$\phi = \phi^\dagger$, which is  a self-adjoint operator acting
on the representation space $\cH$ of the algebra \eq{CCR}.
The action is chosen to be written with an imaginary coupling $i\tilde \la$, 
assuming $\tilde \la$ to be real. The reason is that for real coupling
$\tilde \la' = i\tilde \la$, the potential 
would be unbounded from above and below, 
and the quantization would seem ill-defined. The quantization is 
completely well-defined for imaginary 
$i\tilde \la$, 
and allows analytic continuation to real $\tilde \la' = i\tilde \la$
in a certain sense which will be made precise below.
Therefore we accept for now that the action $\tilde S$ is not
necessarily  real.
Using the commutation relations \eq{CCR}, the derivatives $\partial_i$
can be written as inner derivatives
$\partial_i f = -i[\tilde x_i,f]$.
Therefore the action can be written as 
\be 
\tilde S = \int -(\tilde
x_i\phi\tilde x_i\phi - \tilde x_i \tilde x_i \phi\phi) + \Omega^2
\tilde x_i \phi \tilde x_i \phi + \frac {\mu^2}2 \phi^2 +
\frac{i\tilde \lambda}{3!}\;\phi^3 
\ee 
using the cyclic property of the
integral. 
For the ``self-dual'' point $\Omega =1$, this action simplifies
further to 
\be
\tilde S = \int (\tilde x_i \tilde x_i + \frac
{\mu^2}2) \phi^2 + \frac{i\tilde \lambda}{3!}\;\phi^3 \,
=\, Tr\Big(\frac 1{2} J \phi^2 +
 \frac{i\lambda}{3!}\;\phi^3 \Big).
\label{action-E} 
\ee
Here we replaced the integral by $\int = (2\pi \theta)^2 Tr$, and
introduce 
\be
J = 2(2\pi \theta)^2 (\sum_i \tilde x_i \tilde x_i + \frac
{\mu^2}2 ),\qquad
\la = (2\pi \theta)^2\tilde \lambda.
\label{const-defs}
\ee 

In \cite{Grosse:2005ig,Grosse:2006qv} it has been shown that
noncommutative Euclidean selfdual $\phi^3$ model
can be solved using matrix model techniques, 
and is related to the KdV hierarchy. 
This is achieved by rewriting 
the field theory as Kontsevich matrix model, for a suitable choice
of the eigenvalues in the latter. The relation holds 
for any even dimension, and allows to apply some of the known, remarkable 
results for the Kontsevich model to the quantization of the 
$\phi^3$ model \cite{Itzykson:1992ya,Kontsevich:1992ti}.  

In order to quantize the theory,  we need to include a linear
counterterm $-Tr (i \la) a\, \phi$ to the action (the explicit factor
$i\la$ is inserted to keep most quantities real), and -- 
as opposed to the 2-dimensional case \cite{Grosse:2005ig} -- 
we must also allow for a divergent
shift 
\be
\phi \to \phi + i\la c
\label{phi-shift}
\ee
of the field $\phi$. 
These counterterms are necessary to ensure that the local 
minimum of the cubic potential remains at the origin after quantization.
The latter shift implies in particular 
that the linear counterterm 
picks up a contribution
$-Tr (i\la)(a+ c J)\phi $ 
from the quadratic term.
Therefore the linear term should be replaced by $-Tr (i\la) A\phi$
where
\be
A = a+c J,
\label{A-def}
\ee
while the other effects of this shift $\phi \to \phi + i \la c$ 
can be absorbed by
a redefinition of the coupling constants 
(which we do not keep track of). 
We are thus led to consider the action
\be 
S = \,Tr \Big( \frac 12 J
\phi^2 + \frac{i \la}{3!}\;\phi^3 - (i\la) A \phi 
- \frac 1{3(i\la)^2}
J^3 -  J A\Big).
\label{action-kontsevich}
\ee 
involving the constants $i\la, \, a,\, c$ and $\mu^2$.
The additional constant terms in \eq{action-kontsevich} 
are introduced for later convenience.
By suitable shifts in the field $\phi$, one can now either eliminate
the linear term or the quadratic term in the action,
\be 
S= Tr \Big( -\frac 1{2 i \la} M^2 \tilde\phi + \frac{i
\la}{3!}\;\tilde\phi^3 \Big) 
= \,Tr \Big(\frac 12 M X^2 + \frac{i\la}{3!}\;X^3 -
\frac 1{3(i\la)^2} M^3 \Big)
\label{action-kontsevich-new}
\ee
where\footnote{for the
  quantization, the integral
for the diagonal elements is then defined via analytical continuation,
and the off-diagonal elements remain hermitian since $J$ is diagonal.}
\be 
\tilde\phi = \phi + \frac 1{i\la} J \, = \, X + \frac 1{i\la} M 
\ee 
and
\bea
M &=& \sqrt{J^2 + 2 (i\la)^2 A} 
= \sqrt{\tilde J^2 + 2 (i\la)^2 a-(i\la)^4 c^2} \label{M-def}\\
\tilde J &=& J + (i\la)^2 c.
\eea 
This has precisely
the form of the Kontsevich model \cite{Kontsevich:1992ti}.

The quantization of the model \eq{action-kontsevich} 
resp. \eq{action-kontsevich-new}
is defined by an integral over
all Hermitian $N^2\times N^2$ matrices $\phi$, where $N$ serves as a UV
cutoff. The partition function is defined as
\be 
Z(M) = \int D\tilde \phi \,\exp(- Tr \Big( -\frac 1{2 i \la} M^2
\tilde\phi 
+ \frac{i\la}{3!}\;\tilde\phi^3 \Big)) = e^{F(M)},
\label{Z-again}
\ee 
which is a function of the eigenvalues of $M$ resp. $\tilde J$.
Since $N$ is finite, we can freely switch between the various 
parametrizations \eq{action-kontsevich}, \eq{action-kontsevich-new}
involving $M$, $J$, $\phi$, or $\tilde\phi$.
Correlators 
or ``$n$-point functions'' are defined through
\be
\langle \phi_{i_1 j_1} ...  \phi_{i_n j_n}\rangle
= \frac 1Z\, \int D \phi \,\exp(- S)\,
\phi_{i_1 j_1} ....  \phi_{i_n j_n},
\label{correl-def}
\ee
keeping in mind that each $i_n$ denotes a double-index \cite{Grosse:2006qv}.

This allows to write down closed 
expressions for the genus expansion of the free energy, 
and also for some $n$-point functions by taking derivatives and
using the equations of motion. 
It turns out that the required renormalization 
is determined by the genus $0$ sector only, 
and can be computed explicitly. As for the renormalization procedure, see
\cite{Grosse:2005ig,Grosse:2006qv,Grosse:2006hs}.
All contributions in a genus expansion of any $n$-point function
correlation function are finite and well-defined for finite coupling.  
This implies but is stronger than perturbative renormalization.
One thus obtains
fully renormalized models with nontrivial interaction
which are free of IR/UV diseases. 
All this shows that even though the $\phi^3$ may appear 
ill-defined at first, it is in fact much better under control
than other models.



\section{\label{introduction}Induced gauge theory}

Since elementary particles are most successfully described
by gauge theories it is a big challenge to formulate
consistent gauge theories on non-commutative spaces. Let $u$ be a
unitary element of the algebra such that the scalar fields $\phi$ transform
covariantly: 
\be 
\label{i.1} \phi \mapsto u^* \star \phi \star u,
\,\, u\in \mathcal G. 
\ee 
For a purpose which will become clear in
the sequel, we rewrite the action~(\ref{action}) using $\partial_\mu f = -i[\tilde
x_\mu, f]_\star$ and obtain \bea S_0 = \int d^Dx &\left( \frac{1}{2}
\phi \star [\tilde x_\nu, \,  [\tilde x^\nu, \phi]_\star]_\star +
\frac{\Omega^2}{2} \phi \star \{ \tilde x^\nu , \{ \tilde x_\nu
,\phi\}_\star \}_\star \right. \nonumber
\\
& \left. + \frac{\mu^2}{2} \phi \star \phi
+ \frac{\lambda}{4!} \phi\star \phi  \star \phi \star \phi\right)(x)\;.
\label{action1}
\eea
The approach employed here makes use of two basic ideas.
First, it is well known that the $\star$-multiplication of a coordinate - and also of a function, of
course - with a field is not a covariant process. The product $x^\mu \star \phi$ will not transform
covariantly,
$$
x^\mu \star \phi \nrightarrow u^* \star x^\mu \star \phi \star u\;.
$$
Functions of the coordinates are not effected by the gauge group. The matter field $\phi$ is taken to be
an element of a left module \cite{Jurco:2000fs}. The introduction of covariant coordinates
\be
B_\nu=\tilde x_\nu + A_\nu
\ee
finds a remedy to this situation \cite{Madore:2000en}. The gauge field $A_\mu$ and hence the
covariant coordinates transform in the
following way:
\bea
\label{i.2}
A_\mu & \mapsto & \mathrm{i} u^* \star \partial_\mu u  + u^* \star A_\mu
\star u \,, \\
\nonumber
B_\mu & \mapsto & u^* \star B_\mu \star u
\; .
\eea
Using covariant coordinates we can construct an action invariant under gauge transformations. This
action defines the model for which we shall study the heat kernel expansion:
\bea
S & = & \int d^Dx \left(
\frac{1}{2} \phi \star [B_\nu,\, [B^\nu,\, \phi]_\star ]_\star
+ \frac{\Omega^2}{2}
\phi \star \{B^\nu , \{ B_\nu ,\phi\}_\star \}_\star
\right.
\nonumber
\\
&&
+ \left. \frac{\mu^2}{2} \phi \star \phi
+ \frac{\lambda}{4!} \phi\star \phi  \star \phi \star \phi\right)(x)\;.
\label{gauge-action}
\eea

Secondly, we apply the heat kernel formalism. The gauge field $A_\mu$ is an external, classical gauge
field coupled to $\phi$. In the following sections, we will explicitly
calculate the divergent terms of the one-loop effective action. In the classical case, the divergent
terms determine the dynamics of the gauge field
\cite{Chamseddine:1996zu,Langmann:2001cv,Vassilevich:2003xt}.
There have already been attempts to generalise this approach to the non-commutative realm; for
non-commutative $\phi^4$ theory see \cite{Gayral:2004cs,Gayral:2004cu}.
First steps towards gauge kinetic models have been done in
\cite{Vassilevich:2003yz,Gayral:2004ww,Vassilevich:2005vk}. However, the results there are not completely comparable,
since we have modified the free action and expand around $-\nabla^2 + \Omega^2 \tilde x^2$ rather than $-\nabla^2$.
In addition, some assumptions are not met. In \cite{Vassilevich:2005vk}, the author starts with an action of the form
\be
S'=\int d^D x \, \phi \, (-\nabla^2 + E) \phi,
\ee
where
\beb
& \nabla_\mu = \partial_\mu + L(\lambda_\mu) + R(\rho_\mu),\\
& E= L(l_1) + R(r_1) + L(l_2)\circ R(r_2).
\eeb
$R$ and $L$ denote right and left $\star$-multiplication, respectively. Comparison with our action yields the following
identifications:
\bea
\nonumber
\lambda_\mu & = & -i A_\mu, \\
\rho_\mu & = & i A_\mu,  \nonumber\\
r_1=l_1 & = & \Omega^2 (B_\mu \star B^\mu-\tilde x^2) , \\
r_{2\mu} = l_{2\mu} & = & \sqrt{2} \Omega B_\mu\; . \nonumber
\eea
All fields are assumed to fall off at infinity faster than any power. But $A$ and $B=\tilde x +A$ cannot both drop off
to zero at infinity.

As we will see, the employed method is not manifest gauge invariant. But in the end, various terms add up
to give gauge invariant results.
In this paper we will discuss the case $\Omega=1$ in $D=2$ and $4$ dimensions and the case $\Omega\ne 1$
in $D=2$ dimensions, respectively.


\subsection{\label{model}The model}

Let us start from the action~(\ref{action})
\beb
S & = & \int d^Dx \left(
\frac{1}{2} \phi \star [B_\nu,\, [B^\nu,\, \phi]_\star ]_\star
+ \frac{\Omega^2}{2}
\phi \star \{B^\nu , \{ B_\nu ,\phi\}_\star \}_\star
\right.
\nonumber
\\
&&
+ \left. \frac{\mu^2}{2} \phi \star \phi
+ \frac{\lambda}{4!} \phi\star \phi  \star \phi \star \phi\right)(x)\;.
\eeb
The expansion of $S$ yields
\bea
S & = & S_0 + \int d^Dx\,\frac{1}{2} \phi \star
 \Big(
2 \mathrm{i}  A^\nu \star \partial_\nu \phi
- 2\mathrm{i} \partial_\nu \phi \star A^\nu
\nonumber
\\
&& + 2 (1+\Omega^2) A_\nu \star A^\nu \star \phi
- 2 (1-\Omega^2) A_\nu \star \phi \star A^\nu
\nonumber
\\
&& + 2 \Omega^2 \{ (\theta^{-1} x)_\nu , (A^\nu \star \phi + \phi
\star A^\nu) \}_\star  \Big)\;, \eea where $S_0$ denotes the
quadratic part ot the action. We expand the fields in the matrix
base of the Moyal plane, \be A^\nu(x) =\sum_{p,q \in
\mathbb{N}^{D/2}} A^\nu_{pq} f_{pq}(x)\;, \phi(x) = \sum_{p,q \in
\mathbb{N}^{D/2}} \phi_{pq} f_{pq}(x)\;, \psi(x) = \sum_{p,q \in
\mathbb{N}^{D/2}} \psi_{pq} f_{pq}(x)\;. \ee This choice of basis
simplifies the calculations. In the end, we will again represent the
results in the $x$-basis. Usefull properties of this basis (which we
also use in the Appendix) are reviewed in the appendix of
\cite{Grosse:2003nw}. In order to calculate the effective 1-loop action we need 
to compute the second derivative of the action first. The second derivative of the 
action yields 
\bea
\frac{\delta^2 S}{ \delta \phi^2}(f_{mn})(x)
&=& \sum_{r,s \in \mathbb{N}^2} G_{rs;mn} f_{sr}(x)
\nonumber
\\
&& \hspace{-1.5cm}
+ \sum_{r \in \mathbb{N}^2} \Big(
\frac{\lambda}{3!} \phi \star \phi
+ (1+\Omega^2)\big( B_\nu\star B^\nu -\tilde x^2 \big) \Big)_{rm} f_{rn}(x)
\nonumber
\\
&&\hspace{-1.5cm}
+ \sum_{s \in \mathbb{N}^2} \Big(
\frac{\lambda}{3!} \phi \star \phi
+ (1+\Omega^2) \big(B_\nu \star B^\nu - \tilde x^2 \big) \Big)_{ns} f_{ms}(x)
\nonumber
\\
&& \hspace{-1.5cm}
+ \sum_{r,s \in \mathbb{N}^2}
\Big(\frac{\lambda}{3!} \phi_{rm} \phi_{ns}
- 2 (1-\Omega^2) A_{\nu,rm} A^\nu_{ns} \Big) f_{rs}(x)
\nonumber
\\
&&\hspace{-1.5cm}
+ (1-\Omega^2)\mathrm{i}  \sqrt{\frac{2}{\theta}}
\sum_{r \in \mathbb{N}^2} \Big(
\sqrt{n^1} A^{(1+)}_{\stackrel{r^1}{r^2}\stackrel{m^1}{m^2}}
f_{\stackrel{r^1}{r^2}\stackrel{n^1-1}{n^2}}
- \sqrt{n^1+1}A^{(1-)}_{\stackrel{r^1}{r^2}\stackrel{m^1}{m^2}}
f_{\stackrel{r^1}{r^2}\stackrel{n^1+1}{n^2}}
\nonumber
\\
&&  \hspace*{.5cm}
+ \sqrt{n^2} A^{(2+)}_{\stackrel{r^1}{r^2}\stackrel{m^1}{m^2}}
f_{\stackrel{r^1}{r^2}\stackrel{n^1}{n^2-1}}
- \sqrt{n^2+1} A^{(2-)}_{\stackrel{r^1}{r^2}\stackrel{m^1}{m^2}}
f_{\stackrel{r^1}{r^2}\stackrel{n^1}{n^2+1}}\Big)
\nonumber
\\
&& \hspace{-1.5cm}
- (1-\Omega^2)\mathrm{i} \sqrt{\frac{2}{\theta}}
\sum_{s \in \mathbb{N}^2} \Big(
- \sqrt{m^1+1} A^{(1+)}_{\stackrel{n^1}{n^2}\stackrel{s^1}{s^2}}
f_{\stackrel{m^1+1}{m^2}\stackrel{s^1}{s^2}}
+ \sqrt{m^1} A^{(1-)}_{\stackrel{n^1}{n^2}\stackrel{s^1}{s^2}}
f_{\stackrel{m^1-1}{m^2}\stackrel{s^1}{s^2}}
\nonumber
\\
&& \hspace*{.5cm}
- \sqrt{m^2+1} A^{(2+)}_{\stackrel{n^1}{n^2}\stackrel{s^1}{s^2}}
f_{\stackrel{m^1}{m^2+1}\stackrel{s^1}{s^2}}
+ \sqrt{m^2} A^{(2-)}_{\stackrel{n^1}{n^2}\stackrel{s^1}{s^2}}
f_{\stackrel{m^1}{m^2-1}\stackrel{s^1}{s^2}} \Big)\;,
\label{spp}
\eea
where
\be
A^{(1\pm)}= A^1\pm \mathrm{i} A^2\;,\qquad
A^{(2\pm)}= A^3\pm \mathrm{i} A^4\;.
\ee

We extract the $lk$-component of (\ref{spp}):
\be
\frac{\theta}{2}
\left(\frac{\delta^2 S}{ \delta \phi^2}(f_{mn})\right)_{lk} =
H^0_{kl;mn} + \frac{\theta}{2}B_{kl;mn} \equiv H_{kl;mn}\;,
\label{SHB}
\ee
where
$H^0_{mn;kl}=\frac{\theta}2 \Delta_{mn;kl}$ given in Eq.~(\ref{Gm}).
is the field independent part and
\begin{align}
V_{kl;mn}
&= \Big(
\frac{\lambda}{3!} \phi \star \phi
+ (1+\Omega^2)\big(B_\nu \star B^\nu -\tilde x^2 \big) \Big)_{lm} \delta_{nk}
\nonumber
\\*
&+ \Big(
\frac{\lambda}{3!} \phi \star \phi
+ (1+\Omega^2) \big(B_\nu \star B^\nu -\tilde x^2 \big) \Big)_{nk} \delta_{ml}
\nonumber
\\
& +
\Big(\frac{\lambda}{3!} \phi_{lm} \phi_{nk}
- 2 (1-\Omega^2) A_{\nu,lm} A^\nu_{nk} \Big)
\nonumber
\\
&+ (1-\Omega^2)\mathrm{i}  \sqrt{\frac{2}{\theta}} \Big(
\sqrt{n^1} A^{(1+)}_{\stackrel{l^1}{l^2}\stackrel{m^1}{m^2}}
\delta_{\stackrel{k^1}{k^2}\stackrel{n^1-1}{n^2}}
- \sqrt{n^1+1}A^{(1-)}_{\stackrel{l^1}{l^2}\stackrel{m^1}{m^2}}
\delta_{\stackrel{k^1}{k^2}\stackrel{n^1+1}{n^2}}
\nonumber
\\*
&  \hspace*{7em}
+ \sqrt{n^2} A^{(2+)}_{\stackrel{l^1}{l^2}\stackrel{m^1}{m^2}}
\delta_{\stackrel{k^1}{k^2}\stackrel{n^1}{n^2-1}}
- \sqrt{n^2+1} A^{(2-)}_{\stackrel{l^1}{l^2}\stackrel{m^1}{m^2}}
\delta_{\stackrel{k^1}{k^2}\stackrel{n^1}{n^2+1}}\Big)
\nonumber
\\
&
- (1-\Omega^2)\mathrm{i} \sqrt{\frac{2}{\theta}} \Big(
- \sqrt{m^1+1} A^{(1+)}_{\stackrel{n^1}{n^2}\stackrel{k^1}{k^2}}
\delta_{\stackrel{m^1+1}{m^2}\stackrel{l^1}{l^2}}
+ \sqrt{m^1} A^{(1-)}_{\stackrel{n^1}{n^2}\stackrel{k^1}{k^2}}
\delta_{\stackrel{m^1-1}{m^2}\stackrel{l^1}{l^2}}
\nonumber
\\*
& \hspace*{7em}
- \sqrt{m^2+1} A^{(2+)}_{\stackrel{n^1}{n^2}\stackrel{k^1}{k^2}}
\delta_{\stackrel{m^1}{m^2+1}\stackrel{l^1}{l^2}}
+ \sqrt{m^2} A^{(2-)}_{\stackrel{n^1}{n^2}\stackrel{k^1}{k^2}}
\delta_{\stackrel{m^1}{m^2-1}\stackrel{l^1}{l^2}} \Big)\;.
\label{sppc}
\end{align}
According to \cite{Gayral:2004cs}, the third line of the above equation corresponds to
non-planar contributions.
The above expressions for even dimensions other than four can easily be extracted.

In (\ref{SHB}), we use two different notations for the index
assignment. Given an operator $P$ on the algebra, we write
\be
P f_{mn} = \sum_{k,l} (P_{mn})_{lk} f_{lk} = \sum_{k,l}
f_{lk} P_{kl;mn}\;.
\ee
Then, the composition of two such operators $P,Q$ reads
\bea
P Q f_{mn} &= \sum_{k,l} (Q_{mn})_{lk} (P f_{lk})
= \sum_{k,l,r,s} (Q_{mn})_{lk} (P_{lk})_{rs} f_{rs}
\nonumber
\\*
& = \sum_{k,l} (P f_{lk}) Q_{kl;mn}
= \sum_{k,l,r,s} f_{rs} P_{sr;lk} Q_{kl;mn}\;,
\eea
hence
\be
 [P Q]_{sr;mn}=\sum_{k,l} P_{sr;lk} Q_{kl;mn}\;.
\ee
The trace of such an operator is then given by
\be
\mathrm{Tr} P = \sum_{m,n} P_{mn;nm}\;.
\label{TrP}
\ee

The regulatised one-loop effective action for the model defined by
the classical action (\ref{gauge-action}) is given by \be
\Gamma^\epsilon_{1l}[\phi] = -\frac{1}{2} \int_\epsilon^\infty
\frac{dt}{t} \,\mathrm{Tr}\left( e^{-t H} - e^{-t H^0} \right) \;.
\label{Gamma-e} \ee Using the Duhamel formula, we obtain the
following expansioin:
\bea \Gamma^\epsilon_{1l}[\phi] & = &
-\frac{1}{2} \int_\epsilon^\infty \frac{dt}{t} \,\mathrm{Tr}\left( -
\frac{\theta}{2} t V e^{-t H^0} + \frac{\theta^2}{4} \int_0^t dt' t'
V e^{-t' H^0} V e^{-(t-t') H^0} +\dots\right)
\label{Duhamel-2}\\
\nonumber
& = & \Gamma^\epsilon_{1l,1}[\phi] + \Gamma^\epsilon_{1l,2}[\phi] + \mathcal O(\theta^3)\,.
\eea

The heat kernel $e^{-tH_0}$ of the Schr\"odinger operator (\ref{Gm})
can be calculated from the propagator given in Eq.~(\ref{prop}). In the matrix base
of the Moyal plane, it has the following representation:
\bea
\left( e^{-tH^0}\right)_{mn;kl} & = & e^{-t(\mu^2\theta/2+\Omega D)}
\delta_{m+k,n+l} \prod_{i=1}^{D/2} K_{m^in^i;k^il^i}(t)\;,
\\
K_{m,m+\alpha;l+\alpha,l}(t) & = & \sum_{u=0}^{\textrm {min}(m,l)}
\sqrt{\binom{m}{u}\binom{l}{u}
  \binom{\alpha+m}{m-u}\binom{\alpha+l}{l-u}}
\nonumber
\\
&& \times
\frac{e^{-4\Omega t(\frac{1}{2} \alpha + u)}
(1-e^{-4\Omega t})^{m+l-2u}}{
(1-\frac{(1-\Omega)^2}{(1+\Omega)^2} e^{-4\Omega t})^{\alpha +m+l+1}}
\Big(\frac{4\Omega}{(1+\Omega)^2}\Big)^{\alpha+2u+1}
\Big(\frac{1-\Omega}{1+\Omega}\Big)^{m+l-2u}
\label{comp}\\
& = & \sum_{u=0}^{\textrm {min}(m,l)}
\sqrt{\binom{m}{u}\binom{l}{u}
  \binom{\alpha+m}{m-u}\binom{\alpha+l}{l-u}} \\
\nonumber
& &\times \,
e^{2 \Omega t} \left( \frac{1-\Omega^2}{2\Omega}\sinh (2\Omega t) \right)^{m+l-2u}
X_\Omega(t)^{\alpha+m+l+1}
\;,
\eea
where we have used the definition
\begin{align}
\label{def1}
X_\Omega(t)= \frac{4\Omega}{
(1+\Omega)^2e^{2\Omega t}-(1-\Omega)^2 e^{-2\Omega t}} \; .
\end{align}

For $\Omega=1$,
the interaction part of the action  simplifies,
\bea
V_{kl;mn}
& = & \Big(
\frac{\lambda}{3!} \phi \star \phi
+ 2\big( B_\mu \star B^\mu - \tilde x^2
\big) \Big)_{lm} \delta_{nk}
\nonumber
\\*
&& + \Big(
\frac{\lambda}{3!} \phi \star \phi
+ 2 \big( B_\mu \star B^\mu - \tilde x^2
\big) \Big)_{nk} \delta_{ml}
+ \frac{\lambda}{3!} \phi_{lm} \phi_{nk}
\label{om1}\\
& \equiv & a_{lm} \delta_{nk} + a_{nk}\delta_{ml} + \frac{\lambda }{3!} \phi_{lm}
\phi_{nk}\, ,
\eea
and for the heat kernel we obtain the following simple expression:
\bea
\left( e^{-tH^0} \right)_{mn;kl} & = & \delta_{ml} \delta_{kn} e^{-2t\sigma^2}
\prod_{i=1}^{D/2} e^{-2t(m^i + n^i)},\\
K_{mn;kl}(t) & = & \delta_{ml} \prod_{i=1}^{D/2}e^{-2t(m^i+k^i)},
\eea
where $\sigma^2=\frac{\mu^2\theta}4+\frac{D}2$.


\subsection{\label{d2}2-Dimensional case}


\subsubsection{\label{sec3.1} $\Omega=1$}

The heat kernel is given by
\be
\left( e^{-tH^0} \right)_{mn;kl} = \delta_{ml} \delta_{kn} e^{-2t\sigma^2}
e^{-2t(m + n)},
\ee
where $\sigma^2=\frac{\mu^2\theta}4+1$.
Let us compute the first term in (\ref{Duhamel-2}). In order to do so we need to calculate the partial
trace
\bea
\label{d2.1}
\sum_{n=0}^\infty \left( e^{-tH^0} \right)_{mn;nl} & = &
\sum_{n=0}^\infty e^{-2t\sigma^2} e^{-2t(m+n)} \delta_{ml}
\\
\nonumber
& = & e^{-2t(\sigma^2+m)} \frac1{1-e^{-2t}} \delta_{ml}\\
& \approx & e^{-2t\sigma^2} (\frac1{2t} + \frac12 -m)\delta_{ml} + \mathcal O(t)\; .
\label{d2.2}
\eea
Therefore, we obtain for the divergent contribution of the effective action
\bea
\label{d2.3}
\Gamma^\epsilon_{1l,1} & = & \frac{\theta}{4} \int_\epsilon^\infty dt \,
\mathrm{Tr}(V e^{-t H^0})
\nonumber
\\
\label{d2.4}
& = & \frac{-\theta}{4} \sum_m \Big(
\frac{\lambda}{3!} \phi \star \phi
+  2 \big( B_\mu \star B^\mu - \tilde x^2 \big) \Big)_{mm} \ln \epsilon
\\
\nonumber
&& + \int_\epsilon ^\infty dt\, \frac{\theta}4
\frac{\lambda}{3!} \sum_{m,n} \phi_{mm} \phi_{nn} e^{-2t(\sigma^2+m+n)}
+ \textrm{ finite terms }\, .
\eea
The last term
$$
\int_\epsilon ^\infty dt\, \frac{\theta}4
\frac{\lambda}{3!} \sum_{m,n} \phi_{mm} \phi_{nn} e^{-2t(\sigma^2+m+n)} =
\int_\epsilon ^\infty dt\, \frac{\theta}4\frac{\lambda}{3!} \,  e^{-2t\sigma^2} \, \sum_m \phi_{mm} e^{-2tm}
\sum_n \phi_{nn} e^{-2tn}
$$
is finite, if we assume $\phi$ to be a trace class operator, i.e., $|\sum_m \phi_{mm}|<\infty$.
In future, we will skip the remark "+ finite terms" and assuming that only the divergent contributions are of interest.

We can now use $\sum_{m} \psi_{mm} = \frac{1}{2\pi\theta} \int
d^2x \;\psi(x)$, $\psi(x)=\sum_{m,n} \psi_{mn} f_{mn}(x)$, to transform
(\ref{d2.4}) into
\bea
\label{d2.5}
\Gamma^\epsilon_{1l} & = &
\frac{-1}{8 \pi } \int d^2x \Big(  \frac{\lambda}{3!}
\, \phi \star \phi
+ 2(B_\mu \star B^\mu - \tilde x^2) \Big) \ln \epsilon
\, .
\eea
There are no divergencies from the higher order expansion of (\ref{Duhamel-2}), since
the leading contributions are of the form
\be
\int_\epsilon^\infty dt\,  t e^{-2t(\sigma^2 +c)} \frac{1}{2t},
\ee
which is finite in the limit $\epsilon \to 0$.


\subsubsection{\label{sec3.2}$\Omega \ne 1$}

Due to the off-diagonal terms of the potential $V$ in Eq.~(\ref{sppc}), it is not sufficient to calculate
only the partial trace $(e^{-tH^0})_{mn;nl}$ in order to obtain the first order term $\Gamma^\epsilon_{1l,1}$ of
the effective action. It reads
\bea
\nonumber
\Gamma^\epsilon_{1l,1}[\phi] & = & \frac{\theta}4 \int dt \, \textrm{Tr} (Ve^{-tH^0})\\
& =  & 2 a_{ml} \left( e^{-tH^0} \right)_{mn;nl} + (\frac{\lambda}{3!}\phi_{lm} \phi_{nk} -
2(1-\Omega^2) A_{\nu,lm} A^\nu_{nk}) \left( e^{-tH^0} \right)_{nm;lk}\\
\nonumber
&& + i \sqrt{2/\theta} \left\{ \sqrt{n+1} A^+_{l,m+1} - \sqrt{n+1}A^-_{l,m+1}
            + \sqrt{m+1} A^+_{n+1,l} \right. \\
\nonumber
&& \hspace{3cm}
\left. - \sqrt{m+1} A^-_{n+1,l} \right\}
            \left( e^{-tH^0} \right)_{n+1,m+1;ml}
\eea
As above, we neglect the non-planar contribution. We are left with two different partial traces. For the first one we obtain
the following expression:
\bea
&&
\sum_{n=0}^\infty K_{mn;nm}(t) =
\nonumber
\\
& = &  \sum_{n=0}^\infty \sum_{v=0}^{\min(m,n)}
\binom{m}{v} \binom{n}{v}
\frac{e^{-4\Omega t(\frac{1}{2}n + \frac{1}{2} m - v)}
(1-e^{-4\Omega t})^{2v}}{
(1-\frac{(1-\Omega)^2}{(1+\Omega)^2} e^{-4\Omega t})^{n+m+1}}
\Big(\frac{4\Omega}{(1+\Omega)^2}\Big)^{n+m-2v+1}
\Big(\frac{1-\Omega}{1+\Omega}\Big)^{2v} \,.
\eea
In the limit $t\to 0$ this is equivalent to
\bea
\nonumber
\sum_{n=0}^\infty K_{mn;nm}(t) & \cong &
\sum_{n=0}^\infty \left( \frac{4\Omega}{(1+\Omega)^2 e^{2\Omega t} - (1-\Omega)^2 e^{-2\Omega t}} \right)^n\\
\label{K1}
& = & \frac1{(1+\Omega^2)t} +\mathcal O(1)\,.
\eea
The second expression we need is the sum
\bea
&& \hspace{-2cm}
\sum_{n=0}^\infty  \frac{\sqrt{n+1}}{\sqrt{m+1}}
K_{m+1,n+1;n,m}(t) =
\nonumber
\\
&=& \sum_{n=0}^\infty  \sum_{v=0}^{\min(m,n)}
\frac{\sqrt{n+1}}{\sqrt{m+1}} \sqrt{\binom{m+1}{v+1}\binom{m}{v}
  \binom{n+1}{v+1}\binom{n}{v}}
\nonumber
\\
&& \times
\frac{e^{-2\Omega t(m+n-2v)}
(1-e^{-4\Omega t})^{2v+1}}{
(1-\frac{(1-\Omega)^2}{(1+\Omega)^2} e^{-4\Omega t})^{m +n+2}}
\Big(\frac{4\Omega}{(1+\Omega)^2}\Big)^{m+n-2v+1}
\Big(\frac{1-\Omega}{1+\Omega}\Big)^{2v+1}
\nonumber
\\
\label{K2}
& = & \frac{1-\Omega^2}{(1+\Omega^2)^2} \, \frac1{t} + \mathcal O(1)\,.
\eea

We now obtain the quadratically divergent part of the effective action
by inserting (\ref{K1}) and (\ref{K2})
into (\ref{Gamma-e}), restricted to the first term in
(\ref{Duhamel-2}). With (\ref{sppc}) we have
\begin{align}
\Gamma^\epsilon_{1l}
&=\frac{\theta}{4} \int_\epsilon^\infty dt \,
\mathrm{Tr}(V e^{-t H^0})
\nonumber
\\
&=\frac{-\theta}{2(1+\Omega^2)} \sum_m \Big(
\frac{\lambda}{3!} \phi \star \phi
+ (1+\Omega^2) A_\nu \star A^\nu \Big)_{mm} \ln \epsilon
\nonumber
\\
& - \frac{4 \Omega^2 \theta }{(1+\Omega^2)^2} \sum_m
\Big((\theta^{-1}x)_\nu \cdot A^\nu\Big)_{mm} \ln \epsilon
\label{TrB} \;.
\end{align}

We can now use $\sum_{m} \psi_{mm} = \frac{1}{2\pi\theta} \int
d^2x \;\psi(x)$, $\psi(x)=\sum_{m,n} \psi_{mn} f_{mn}(x)$, to transform
(\ref{TrB}) into
\begin{align}
\Gamma^\epsilon_{1l} &=
\frac{-1}{2\pi (1+\Omega^2)} \int d^2x \Bigg\{
\frac{\lambda}{3!2}
\, \phi \star \phi
+ \frac{1+\Omega^2}{2}
\, A_\nu \star A^\nu
\label{dd1}
\\
& \qquad + \frac{4 \Omega^2}{1+\Omega^2}
\,
(\theta^{-1}x)_\nu \cdot A^\nu \Bigg\}\, \ln \epsilon
\nonumber
\end{align}
This expression is not gauge invariant. However, divergent contributions to the effective action in
second order will repair this defect. According to Eq.~(\ref{Gamma-e}), the action to second order
is given by
\beb
- \frac{\theta^2}{8} \int_\epsilon^\infty \frac{dt}{t} \int_0^t
dt' t' \, V_{mn,kl} \left( e^{-t' H^0} \right)_{lk,sr} V_{rs,vu}
\left( e^{-(t-t') H^0} \right)_{uv,nm} \,.
\eeb
The only divergent contributions are due to off-diagonal elements in the potential $V$.
Explicitly, we obtain
\bea
&& \hspace{-1cm}
- \frac{\theta^2}{8} \int_\epsilon^\infty \frac{dt}{t} \int_0^t
dt' t' \, V_{mn;kl} \left( e^{-t' H^0} \right)_{lk;sr} V_{rs;vu}
\left( e^{-(t-t') H^0} \right)_{uv;nm} =
\nonumber\\
&=&
- \frac{\theta}4 \int_\epsilon^\infty \frac{dt}{t} \int_0^t
dt' t' \, e^{-2t\sigma^2}  (1-\Omega^2)^2 \sqrt{l}
\sqrt{u+1} \, A^+_{nk} A^-_{sv} K_{lk;sr}(t') K_{uv;nm}(t-t')\\
&& \times \delta_{m+1,l}\, \delta_{r,u+1}\, \delta_{l+s,k+r} \,
\delta_{u+n,v+m}
\nonumber\\
&& - \frac{\theta}4 \int_\epsilon^\infty \frac{dt}{t} \int_0^t
dt' t' \, e^{-2t\sigma^2} (1-\Omega^2)^2 \sqrt{l+1}
\sqrt{u} \, A^-_{nk} A^+_{sv} K_{lk;sr}(t') K_{uv;nm}(t-t')\\
&& \times \delta_{m,l+1}\, \delta_{r+1,u}\, \delta_{l+s,k+r} \,
\delta_{u+n,v+m}
\nonumber\\
&& - \frac{\theta}4 \int_\epsilon^\infty \frac{dt}{t} \int_0^t
dt' t' \, e^{-2t\sigma^2} (1-\Omega^2)^2 \sqrt{v}
\sqrt{k+1} \, A^+_{lm} A^-_{ur} K_{lk;sr}(t') K_{uv;nm}(t-t')\\
&& \times \delta_{k+1,n}\, \delta_{v,s+1}\, \delta_{l+s,k+r} \,
\delta_{u+n,v+m}
\nonumber\\
&& - \frac{\theta}4 \int_\epsilon^\infty \frac{dt}{t} \int_0^t
dt' t' \, e^{-2t\sigma^2} (1-\Omega^2)^2 \sqrt{k}
\sqrt{v+1} \, A^-_{lm} A^+_{ur} K_{lk;sr}(t') K_{uv;nm}(t-t')\\
&& \times \delta_{k,n+1}\, \delta_{v+1,s}\, \delta_{l+s,k+r} \,
\delta_{u+n,v+m}
\nonumber
\eea
Each of the 4 terms yields the same contribution. Therefore, let us concentrate
on the first one. In the limit $t\to 0,\, t'\to 0$ this is equivalent to
\bea
\nonumber
&&\hspace{-1cm}
- \frac{\theta}4 \int_\epsilon^\infty \frac{dt}{t} \int_0^t
dt' t' \, e^{-2t\sigma^2}  (1-\Omega^2)^2 (m+1)
\, A^+_{nk} A^-_{kn} K_{m+1,k;k,m+1}(t') K_{mn;nm}(t-t') =\\
\label{sum-m}
&=&- \frac{\theta}4 \int_\epsilon^\infty \frac{dt}{t} \int_0^t
dt' t' \, e^{-2t\sigma^2}  (1-\Omega^2)^2 (m+1)\, A^+_{nk} A^-_{kn}\\
&& \hspace{-.7cm}
\nonumber
\times
\sum_{m=0}^\infty (m+1) \left( \frac{(4\Omega)^2} {((1+\Omega)^2e^{2\Omega t'}-(1-\Omega)^2
e^{-2\Omega t'}) ((1+\Omega)^2e^{2\Omega (t-t')}-(1-\Omega)^2
e^{-2\Omega (t-t')})} \right)^m\\
\nonumber
&=&- \frac{\theta}4 \int_\epsilon^\infty \frac{dt}{t^3} \int_0^t
dt' t' \, e^{-2t\sigma^2}  \frac{(1-\Omega^2)^2}{(1+\Omega^2)^2}
\sum_{nk} A^+_{nk} A^-_{kn}
+\mathcal O(t^0)\\
&=& \frac1{2\pi\theta} \int d^2 x\, \frac{\theta}8 \left(\frac{1-\Omega^2}{1+\Omega^2} \right)^2 A_\mu A^\nu \ln\epsilon
+\mathcal O(t^0)
\eea
Therefore, we obtain the following final result:
\bea
\Gamma^\epsilon_{1l} &=&
\frac{-1}{2\pi (1+\Omega^2)} \int d^2x \Bigg\{
\frac{\lambda}{3!2}
\, \phi \star \phi
+ \frac{2\Omega^2}{1+\Omega^2}
\, ( B_\nu \star B^\nu - \tilde x^2) \Bigg\} \ln\epsilon
\eea


\subsection{\label{d4}4-Dimensional case}

In 4 dimensions the calculations are much more involved. Higher than second order contributions
will contribute to the effective action. Therefore, we restrict
ourselves to the case $\Omega=1$.
We start with the heat kernel in $D=4$ dimensions:
\be
\label{d4.1}
\left( e^{-tH^0} \right)_{mn;kl} = \delta_{ml} \delta_{kn} e^{-2t\sigma^2}
\prod_{i=1}^{2} e^{-2t(m^i + n^i)},
\ee
where $\sigma^2=\frac{\mu^2\theta}4+ 2$.

\bea
\label{d4.2}
\sum_{n=0}^\infty \left( e^{-tH^0} \right)_{mn;nl} & = & \delta_{ml} \sum_{n^1,n^2=0}^\infty e^{-2t\sigma^2} e^{-2t(m^1+n^1)}
    e^{-2t(m^2+n^2)}
\\
\nonumber
& = & e^{-2t(\sigma^2+m^1+m^2)} \delta_{ml} \frac1{(1-e^{-2t})^2} \\
& \approx & e^{-2t(\sigma^2+m)} \frac{\delta_{ml}}{4t^2} (1 - 2t ( m^1+m^2-1 )) + \mathcal O(1) \, .
\label{d4.3}
\eea
Next, let us calculate the divergent contributions of $\Gamma^\epsilon_{1l,1}$, the first term in the expansion of
the 1-loop effective action. In order to do so we have to evaluate the integral over the following trace:
\bea
\Gamma^\epsilon_{1l,1} & = & \frac{\theta}{4} \int_\epsilon^\infty dt \,
\mathrm{Tr}(V e^{-t H^0})
\nonumber
\\
& = & \frac{\theta}{8} \sum_m \Big(
\frac{\lambda}{3!} \phi \star \phi
+ 2\big( B_\mu \star B^\mu - \tilde x^2\big) \Big)_{mm}
\left(\frac{1}{\epsilon} + \frac{\mu^2\theta}{2}
\ln \epsilon \right)
\nonumber
\\
&& + \frac{\theta}{4} \sum_m
(m^1+m^2+1)\Big(
\frac{\lambda}{3!} \phi \star \phi
+ 2\big( B_\mu \star B^\mu - \tilde x^2\big)
 \Big)_{mm}
\ln \epsilon
\nonumber
\;.
\eea
As we have seen before, the contribution of the last term is finite.
At this point, we again perform a change of basis from the matrix basis
to the x-basis using $\sum_m \psi_{mm}= \frac1{(2\pi \theta)^2} \int d^4 x\, \psi(x)$,
where $\psi(x)=\sum_{m,n} \psi_{mn}f_{mn}(x)$.
Using 
\be
\sum_m \left( m^1 + m^2 + 1 \right) (\psi \star \psi)_{mm}
=\frac1{(2\pi\theta)^2} \int d^4 x\, \frac{\theta}8 \psi \star
    \left( - \Delta + 4 \tilde x^2 \right) \psi\,
\ee
we obtain as contributions to the effective action
\bea
\label{d4.10}
\Gamma^\epsilon_{1l,1} & = & \frac1{16\pi^2} \int d^4 x\,
\Bigg(
\frac{\lambda}{3!2\theta}
    \left(\frac1{\epsilon} + \frac{\mu^2 \theta}2 \ln \epsilon \right)
    \phi \star \phi \\
\nonumber
&&
+ \frac{\lambda}{3! 8}  \phi \star (-\Delta + 4\tilde x^2)\phi \, \ln \epsilon
\\
\nonumber
&&
+ \frac1{\theta} \left(\frac1{\epsilon}+\frac{\mu^2\theta}2 \ln \epsilon \right)
    (B_\mu \star B^\mu - \tilde x^2)\\
\nonumber
&&
+ \frac14 A_\mu \star (-\Delta + 4\tilde x^2) A^\mu \ln \epsilon +
    2 \tilde x^2 \tilde x_\mu A^\mu \ln \epsilon
\Bigg),
\eea
where we have used the identification $B_\mu = A_\mu + \tilde x_\mu$.
The third line of Eq.~(\ref{d4.10}) is gauge invariant, but not the fourth one. In order to render
the 1-loop effective action gauge invariant, we need further terms from the
second order expansion of (\ref{Duhamel-2}). But let us first rewrite the that last line using
$\partial_\mu a(x) = -i[\tilde x_\mu, a(x)]_\star$,
\bea
\label{d4.12}
\frac14 A_\mu \star (-\Delta + 4 \tilde x^2)A^\mu + 2\tilde x^2 \tilde x_\mu A^\mu
= \tilde x^2 (A_\mu \star A^\mu) + 2\tilde x^2 \tilde x_\mu A^\mu\,.
\eea
Hence $\Gamma^\epsilon_{1l,1}$ reads
\bea
\label{d4.13}
\Gamma^\epsilon_{1l,1} & = & \frac1{16\pi^2} \int d^4 x\,
\Bigg(
\frac{\lambda}{3!2\theta}
    \left(\frac1{\epsilon} + \frac{\mu^2 \theta}2 \ln \epsilon \right)
    \phi \star \phi \\
\nonumber
&&
+ \frac{\lambda}{3! 8}  \phi \star (-\Delta + 4\tilde x^2)\phi \, \ln \epsilon
\\
\nonumber
&&
+ \frac1{\theta} \left(\frac1{\epsilon}+\frac{\mu^2\theta}2 \ln \epsilon \right)
    (B_\mu \star B^\mu - \tilde x^2)\\
\nonumber
&&
+ \tilde x^2 (A_\mu \star A^\mu) \ln \epsilon + 2\tilde x^2 \tilde x_\mu A^\mu \ln \epsilon
\Bigg)\, .
\eea

The second order expansion of (\ref{Duhamel-2}) has the form
\be
\Gamma^\epsilon_{1l,2}[\phi]=-\frac{\theta^2}{8} \sum_{m,n,k,l,r,s,u,v}
\int_\epsilon^\infty dt t \int_0^1 d\xi \xi
V_{mn;kl} \Big(e^{-t\xi H^0}\Big)_{lk;sr} V_{rs;vu}\Big(e^{-t(1-\xi)
H^0}\Big)_{uv;nm}\;.
\label{BB}
\ee
Let us first consider the case $\phi=0$. Then, we have
\bea
\label{d4.14}
\Gamma^\epsilon_{1l,2}[\phi] & = & -\frac{\theta^2}8 \int_\epsilon^\infty dt\,
    t\int_0^1 d\xi \, \xi \sum (a_{nk} \delta_{ml} + a_{lm}\delta_{nk}) \delta_{lr}
    \delta_{ks} e^{-2t\xi(\sigma^2+l^1+l^2+k^1+k^2)}\\
\nonumber
&&
\times (a_{sv} \delta_{ru} + a_{ur}\delta_{sv}) \delta_{um} \delta_{vn} e^{-2t(1-\xi)(\sigma^2+u^1+u^2+v^1+v^2)}
\\
\nonumber
& = & -\frac{\theta^2}8 \int_\epsilon^\infty dt\, t\int_0^1 d\xi\,
2\xi e^{-2t\sigma^2} \Big( \sum_{n,s,m} a_{ns}a_{sn} e^{-2t\xi (s^1+s^2) - 2t(1-\xi)(n^1+n^2) - 2t(m^1+m^2)}
\\
\label{d4.15}
&& 
+ \sum_{m,n} a_{mm}a_{nn}e^{-2t(m^1+m^2+n^1+n^2)} \Big)\,.
\eea
The first line of Eq.~(\ref{d4.15}) contains the following divergent contribution:
\bea
&&-\frac{\theta^2}4 \int_\epsilon^\infty dt\, t\int_0^1 d\xi\,
\xi e^{-2t\sigma^2} \sum_{n,s,m} a_{ns}a_{sn} e^{-2t\xi (s^1+s^2) - 2t(1-\xi)(n^1+n^2) - 2t(m^1+m^2)} = \\
\nonumber
&&\hspace{1cm}
\approx 
\frac1{16\pi^2} \int d^4 x \Big( \frac12 (B_\mu\star B^\mu)\star (B_\nu\star B^\nu)- \frac12 (\tilde x^2)^2
-\tilde x^2(A_\mu \star A^\mu) - 2\tilde x^2 \tilde x_\mu A^\mu \Big) \ln \epsilon
\eea

Last but not least, we have to collect all the divergent contributions containing the
scalar field $\phi$. As we have seen, the divergent contribution from second order is
proportional to $a^{\star 2}= (\frac{\lambda}{3!}\phi\star\phi +2(B_\mu\star
B^\mu-\tilde x^2))^{\star 2}$:
\bea
\nonumber
\Gamma_{1l,2}^\epsilon
& = & \frac1{16\pi^2} \int d^4x  \frac18 (\frac{\lambda}{3!}\phi\star\phi
+ 2 (B_\mu \star B^\mu - \tilde x^2))^{\star 2} \ln\epsilon
\\
&=& \frac1{16\pi^2} \int d^4x  \frac18 \Bigg( 4 (B_\mu \star B^\mu - \tilde x^2)^{\star 2}
\\
&&
+ \frac{\lambda^2}{36} \phi\star\phi\star\phi\star\phi + \frac{2 \lambda}{3} (A_\mu \star A^\mu + 2\tilde x_\mu A^\mu)
\star\phi\star \phi
\Bigg) \ln \epsilon
\eea
Collecting all the terms together, we get for the divergent contributions of the
effective action
\bea 
\nonumber
\Gamma^\epsilon_{1l,1}+\Gamma^\epsilon_{1l,2} & = &
\frac1{16\pi^2} \int d^4 x\, \Bigg( \frac{\lambda}{3!2\theta}\,
\frac1{\epsilon} \, \phi\star \phi + \frac1{\epsilon\theta}
(B_\nu \star B^\nu -\tilde x^2)\\
\nonumber
&&
+ \frac{\lambda}{3!2} \bigg(
\frac14 \phi\star [B_\nu,[B^\nu, \phi]_\star]_\star +
\frac14 \phi\star \{ B_\nu, \{ B^\nu, \phi \}_\star \}_\star \\
\label{result2}
&&
\hspace{3cm}
+ \frac{\mu^2}2 \phi\star \phi + \frac{\lambda}{4!} \phi\star \phi\star\phi\star \phi
\bigg) \ln\epsilon\\
\nonumber && + \bigg( \frac{\mu^2}2 (B_\nu \star B^\nu -\tilde x^2)
+ \frac12 \left( (B_\mu\star B^\mu)\star (B_\nu\star B^\nu)- (\tilde
x^2)^2 \right) \bigg) \ln \epsilon \Bigg) \eea


\subsection{\label{conclusions}Remarks}

There are three different regimes corresponding to different values of $\Omega$. Namely, for $\Omega=0$, we obtain
the usual non-commutative theories on $\mathbb R_\theta$ with IR/UV mixing catastrophe.
The work presented here implies that for $\Omega=1$ the gauge action consists of a static potential, only.
No dynamical term, such as $F_{\mu\nu} F^{\mu\nu}$ appears. The resulting potentials for $2$ and $4$ dimensions are given in
(\ref{d2.5}) and (\ref{result2}), respectively.

The third regime is $\Omega\ne 0$. The 2 dimensional case has been treated here. Remarkably, first order and second order
contributions add up to give a gauge invariant result. In the case $\Omega=1$, there were no divergent second order contributions
to the effective action. Due to the off-diagonal terms of the potential~$V$ (\ref{sppc}) divergent terms appear, see
Eq.~(\ref{sum-m}).

And
in a next step, we have to study the model away from the selfduality point in 4 dimensions.
Logarithmic divergent dynamical terms, such as $F_{\mu\nu} F^{\mu\nu}$ are supposed to occur. They will appear with
a factor proportional to $1-\Omega^2$. These contributions are therefore absent in the selfdual case.
We intend to study the question of renormalizability for these resulting noncommutative
gauge field models.

\subsubsection*{Acknowledgement}
One of us (H.G.) wants to thank G.~Zoupanos for the invitation to the very
pleasant and stimulating workshop. He also wants to thank R.~Wulkenhaar for the longstanding and
enjoyable collaboration on NCFT and renormalization, and H.~Steinacker for the
fruitful collaboration on the construction of the noncommutative $\phi^3$ model.
\vspace{1cm}

\bibliographystyle{utphys}

\bibliography{tuw}

\providecommand{\href}[2]{#2}\begingroup\raggedright\begin{thebibliography}{10}

\bibitem{Snyder:1947qz}
H.~S. Snyder, ``{Quantized Space-Time},'' {\em Phys. Rev.} {\bf 71} (1947)
38--41.

\bibitem{Connes:1986uh}
A.~Connes, ``Noncommutative differential geometry,'' {\em Inst. Hautes Etudes
  Sci. Publ. Math.} {\bf 62} (1986)
257.

\bibitem{Grosse:1992bm}
H.~Grosse and J.~Madore, ``A noncommutative version of the {S}chwinger model,''
  {\em Phys. Lett.} {\bf B283} (1992)
218--222.

\bibitem{Grosse:1995ar}
H.~Grosse, C.~Klimcik, and P.~Presnajder, ``Towards finite quantum field theory
  in noncommutative geometry,'' {\em Int. J. Theor. Phys.} {\bf 35} (1996)
  231--244,
\href{http://www.arXiv.org/abs/hep-th/9505175}{{\tt hep-th/9505175}}.

\bibitem{Filk:1996dm}
T.~Filk, ``Divergencies in a field theory on quantum space,'' {\em Phys. Lett.}
  {\bf B376} (1996)
53--58.

\bibitem{Doplicher:1994tu}
S.~Doplicher, K.~Fredenhagen, and J.~E. Roberts, ``The quantum structure of
  space-time at the {P}lanck scale and quantum fields,'' {\em Commun. Math.
  Phys.} {\bf 172} (1995) 187--220,
\href{http://www.arXiv.org/abs/hep-th/0303037}{{\tt hep-th/0303037}}.

\bibitem{Schomerus:1999ug}
V.~Schomerus, ``D-branes and deformation quantization,'' {\em JHEP} {\bf 06}
  (1999) 030,
\href{http://www.arXiv.org/abs/hep-th/9903205}{{\tt hep-th/9903205}}.

\bibitem{Seiberg:1999vs}
N.~Seiberg and E.~Witten, ``String theory and noncommutative geometry,'' {\em
  JHEP} {\bf 09} (1999) 032,
\href{http://arXiv.org/abs/hep-th/9908142}{{\tt hep-th/9908142}}.

\bibitem{Szabo:2001kg}
R.~J. Szabo, ``Quantum field theory on noncommutative spaces,'' {\em Phys.
  Rept.} {\bf 378} (2003) 207--299,
\href{http://www.arXiv.org/abs/hep-th/0109162}{{\tt hep-th/0109162}}.

\bibitem{Douglas:2001ba}
M.~R. Douglas and N.~A. Nekrasov, ``Noncommutative field theory,'' {\em Rev.
  Mod. Phys.} {\bf 73} (2002) 977--1029,
\href{http://arXiv.org/abs/hep-th/0106048}{{\tt hep-th/0106048}}.

\bibitem{Minwalla:1999px}
S.~Minwalla, M.~Van~Raamsdonk, and N.~Seiberg, ``Noncommutative perturbative
  dynamics,'' {\em JHEP} {\bf 02} (2000) 020,
\href{http://www.arXiv.org/abs/hep-th/9912072}{{\tt hep-th/9912072}}.

\bibitem{Chepelev:1999tt}
I.~Chepelev and R.~Roiban, ``Renormalization of quantum field theories on
  noncommutative {$\mathbb R^d$}. {I}: Scalars,'' {\em JHEP} {\bf 05} (2000)
  037,
\href{http://www.arXiv.org/abs/hep-th/9911098}{{\tt hep-th/9911098}}.

\bibitem{Chepelev:2000hm}
I.~Chepelev and R.~Roiban, ``Convergence theorem for non-commutative {F}eynman
  graphs and renormalization,'' {\em JHEP} {\bf 03} (2001) 001,
\href{http://www.arXiv.org/abs/hep-th/0008090}{{\tt hep-th/0008090}}.

\bibitem{Grosse:2004yu}
H.~Grosse and R.~Wulkenhaar, ``Renormalisation of {$\phi^4$} theory on
  noncommutative {$\mathbb R^4$} in the matrix base,'' {\em Commun. Math.
  Phys.} {\bf 256} (2005) 305--374,
\href{http://www.arXiv.org/abs/hep-th/0401128}{{\tt hep-th/0401128}}.

\bibitem{Langmann:2002cc}
E.~Langmann and R.~J. Szabo, ``Duality in scalar field theory on noncommutative
  phase spaces,'' {\em Phys. Lett.} {\bf B533} (2002) 168--177,
\href{http://www.arXiv.org/abs/hep-th/0202039}{{\tt hep-th/0202039}}.

\bibitem{Rivasseau:2005bh}
V.~Rivasseau, F.~Vignes-Tourneret, and R.~Wulkenhaar, ``Renormalization of
  noncommutative phi**4-theory by multi- scale analysis,'' {\em Commun. Math.
  Phys.} {\bf 262} (2006) 565--594,
\href{http://www.arXiv.org/abs/hep-th/0501036}{{\tt hep-th/0501036}}.

\bibitem{Grosse:2005ig}
H.~Grosse and H.~Steinacker, ``Renormalization of the noncommutative phi**3
  model through the kontsevich model,'' {\em Nucl. Phys.} {\bf B746} (2006)
  202--226,
\href{http://www.arXiv.org/abs/hep-th/0512203}{{\tt hep-th/0512203}}.

\bibitem{Grosse:2006qv}
H.~Grosse and H.~Steinacker, ``A nontrivial solvable noncommutative $\phi^3$
  model in 4 dimensions,''
\href{http://www.arXiv.org/abs/hep-th/0603052}{{\tt hep-th/0603052}}.

\bibitem{Grosse:2006hs}
H.~Grosse and H.~Steinacker. Work in progress.

\bibitem{Gubser:2000cd}
S.~S. Gubser and S.~L. Sondhi, ``Phase structure of non-commutative scalar
  field theories,'' {\em Nucl. Phys.} {\bf B605} (2001) 395--424,
\href{http://www.arXiv.org/abs/hep-th/0006119}{{\tt hep-th/0006119}}.

\bibitem{Bietenholz:2004xs}
W.~Bietenholz, F.~Hofheinz, and J.~Nishimura, ``Phase diagram and dispersion
  relation of the non- commutative lambda phi**4 model in d = 3,'' {\em JHEP}
  {\bf 06} (2004) 042,
\href{http://www.arXiv.org/abs/hep-th/0404020}{{\tt hep-th/0404020}}.

\bibitem{Martin:2004un}
X.~Martin, ``A matrix phase for the {$\phi^4$} scalar field on the fuzzy
  sphere,'' {\em JHEP} {\bf 04} (2004) 077,
\href{http://www.arXiv.org/abs/hep-th/0402230}{{\tt hep-th/0402230}}.

\bibitem{Ambjorn:2000nb}
J.~Ambjorn, Y.~M. Makeenko, J.~Nishimura, and R.~J. Szabo, ``Nonperturbative
  dynamics of noncommutative gauge theory,'' {\em Phys. Lett.} {\bf B480}
  (2000) 399--408,
\href{http://www.arXiv.org/abs/hep-th/0002158}{{\tt hep-th/0002158}}.

\bibitem{Carow-Watamura:1998jn}
U.~Carow-Watamura and S.~Watamura, ``Noncommutative geometry and gauge theory
  on fuzzy sphere,'' {\em Commun. Math. Phys.} {\bf 212} (2000) 395--413,
\href{http://www.arXiv.org/abs/hep-th/9801195}{{\tt hep-th/9801195}}.

\bibitem{Steinacker:2003sd}
H.~Steinacker, ``Quantized gauge theory on the fuzzy sphere as random matrix
  model,'' {\em Nucl. Phys.} {\bf B679} (2004) 66--98,
\href{http://www.arXiv.org/abs/hep-th/0307075}{{\tt hep-th/0307075}}.

\bibitem{Behr:2005wp}
W.~Behr, F.~Meyer, and H.~Steinacker, ``Gauge theory on fuzzy {$S^2 \times
  S^2$} and regularization on noncommutative {$\mathbb{R}^4$},''
\href{http://www.arXiv.org/abs/hep-th/0503041}{{\tt hep-th/0503041}}.

\bibitem{Gilkey:1995mj}
P.~B. Gilkey, ``Invariance theory, the heat equation and the atiyah-singer
  index theorem,''.

\bibitem{Vassilevich:2003xt}
D.~V. Vassilevich, ``Heat kernel expansion: User's manual,'' {\em Phys. Rept.}
  {\bf 388} (2003) 279--360,
\href{http://www.arXiv.org/abs/hep-th/0306138}{{\tt hep-th/0306138}}.

\bibitem{Grosse:2006mw}
H.~Grosse and M.~Wohlgenannt, ``{Induced Gauge Theory on Non-Commutative
  Spaces},''. Work in progress.

\bibitem{Grosse:2003aj}
H.~Grosse and R.~Wulkenhaar, ``Power-counting theorem for non-local matrix
  models and renormalisation,'' {\em Commun. Math. Phys.} {\bf 254} (2005)
  91--127,
\href{http://www.arXiv.org/abs/hep-th/0305066}{{\tt hep-th/0305066}}.

\bibitem{Polchinski:1983gv}
J.~Polchinski, ``Renormalization and effective {La}grangians,'' {\em Nucl.
  Phys.} {\bf B231} (1984)
269--295.

\bibitem{Wilson:1973jj}
K.~G. Wilson and J.~B. Kogut, ``The renormalization group and the epsilon
  expansion,'' {\em Phys. Rept.} {\bf 12} (1974)
75--200.

\bibitem{Grosse:2004by}
H.~Grosse and R.~Wulkenhaar, ``The beta-function in duality-covariant
  noncommutative {$\phi^4$} theory,'' {\em Eur. Phys. J.} {\bf C35} (2004)
  277--282,
\href{http://www.arXiv.org/abs/hep-th/0402093}{{\tt hep-th/0402093}}.

\bibitem{Rivasseau:1991ub}
V.~Rivasseau, ``From perturbative to constructive renormalization,''.
  Princeton, USA: Univ. Pr. (1991) 336 p. (Princeton series in physics).

\bibitem{Itzykson:1992ya}
C.~Itzykson and J.~B. Zuber, ``Combinatorics of the modular group. 2. the
  {K}ontsevich integrals,'' {\em Int. J. Mod. Phys.} {\bf A7} (1992)
  5661--5705,
\href{http://www.arXiv.org/abs/hep-th/9201001}{{\tt hep-th/9201001}}.

\bibitem{Kontsevich:1992ti}
M.~Kontsevich, ``Intersection theory on the moduli space of curves and the
  matrix {Airy} function,'' {\em Commun. Math. Phys.} {\bf 147} (1992)
1--23.

\bibitem{Jurco:2000fs}
B.~Jur\v{c}o, P.~Schupp, and J.~Wess, ``Noncommutative gauge theory for
  {Poisson} manifolds,'' {\em Nucl. Phys.} {\bf B584} (2000) 784--794,
\href{http://arXiv.org/abs/hep-th/0005005}{{\tt hep-th/0005005}}.

\bibitem{Madore:2000en}
J.~Madore, S.~Schraml, P.~Schupp, and J.~Wess, ``Gauge theory on noncommutative
  spaces,'' {\em Eur. Phys. J.} {\bf C16} (2000) 161--167,
\href{http://www.arXiv.org/abs/hep-th/0001203}{{\tt hep-th/0001203}}.

\bibitem{Chamseddine:1996zu}
A.~H. Chamseddine and A.~Connes, ``The spectral action principle,'' {\em
  Commun. Math. Phys.} {\bf 186} (1997) 731--750,
\href{http://www.arXiv.org/abs/hep-th/9606001}{{\tt hep-th/9606001}}.

\bibitem{Langmann:2001cv}
E.~Langmann, ``Generalized {Yang-Mills actions from Dirac} operator
  determinants,'' {\em J. Math. Phys.} {\bf 42} (2001) 5238--5256,
\href{http://www.arXiv.org/abs/math-ph/0104011}{{\tt math-ph/0104011}}.

\bibitem{Gayral:2004cs}
V.~Gayral, ``Heat-kernel approach to {UV/IR} mixing on isospectral deformation
  manifolds,'' {\em Annales Henri Poincare} {\bf 6} (2005) 991--1023,
\href{http://www.arXiv.org/abs/hep-th/0412233}{{\tt hep-th/0412233}}.

\bibitem{Gayral:2004cu}
V.~Gayral, J.~M. Gracia-Bondia, and F.~R. Ruiz, ``Trouble with space-like
  noncommutative field theory,'' {\em Phys. Lett.} {\bf B610} (2005) 141--146,
\href{http://www.arXiv.org/abs/hep-th/0412235}{{\tt hep-th/0412235}}.

\bibitem{Vassilevich:2003yz}
D.~V. Vassilevich, ``Non-commutative heat kernel,'' {\em Lett. Math. Phys.}
  {\bf 67} (2004) 185--194,
\href{http://www.arXiv.org/abs/hep-th/0310144}{{\tt hep-th/0310144}}.

\bibitem{Gayral:2004ww}
V.~Gayral and B.~Iochum, ``The spectral action for {M}oyal planes,'' {\em J.
  Math. Phys.} {\bf 46} (2005) 043503,
\href{http://www.arXiv.org/abs/hep-th/0402147}{{\tt hep-th/0402147}}.

\bibitem{Vassilevich:2005vk}
D.~V. Vassilevich, ``Heat kernel, effective action and anomalies in
  noncommutative theories,'' {\em JHEP} {\bf 08} (2005) 085,
\href{http://www.arXiv.org/abs/hep-th/0507123}{{\tt hep-th/0507123}}.

\bibitem{Grosse:2003nw}
H.~Grosse and R.~Wulkenhaar, ``Renormalisation of {$\phi^4$} theory on
  noncommutative {$\mathbb R^2$} in the matrix base,'' {\em JHEP} {\bf 12}
  (2003) 019,
\href{http://www.arXiv.org/abs/hep-th/0307017}{{\tt hep-th/0307017}}.

\end{thebibliography}\endgroup

\end{document}